\documentclass[usenatbib,usegraphicx]{mn2e}
\usepackage{amsmath,amssymb}

\def\HI{\hbox{H~$\scriptstyle\rm I\ $}}

\def\HII{\hbox{H~$\scriptstyle\rm II\ $}}

\def\CIII{\hbox{C~$\scriptstyle\rm III\ $}}
\def\CIV{\hbox{C~$\scriptstyle\rm IV\ $}}
\def\NII{\hbox{N~$\scriptstyle\rm II\ $}}
\def\NIII{\hbox{N~$\scriptstyle\rm III\ $}}

\def\OVI{\hbox{O~$\scriptstyle\rm VI\ $}}
\def\SiIII{\hbox{Si~$\scriptstyle\rm III\ $}}
\def\SiIV{\hbox{Si~$\scriptstyle\rm IV\ $}}
\def\SIII{\hbox{S~$\scriptstyle\rm III\ $}}
\def\SIV{\hbox{S~$\scriptstyle\rm IV\ $}}
\def\kms{\,{\rm {km\, s^{-1}}}}
\def\msun{{\rm M_\odot}}

\def\ltsima{$\; \buildrel < \over \sim \;$}
\def\lsim{\lower.5ex\hbox{\ltsima}}
\def\gtsima{$\; \buildrel > \over \sim \;$}
\def\gsim{\lower.5ex\hbox{\gtsima}}

\def\spose#1{\hbox to 0pt{#1\hss}}
\def\lta{\mathrel{\spose{\lower 3pt\hbox{$\mathchar"218$}}
     \raise 2.0pt\hbox{$\mathchar"13C$}}}
\def\gta{\mathrel{\spose{\lower 3pt\hbox{$\mathchar"218$}}
     \raise 2.0pt\hbox{$\mathchar"13E$}}}

\journal{Preprint-00}

\title{Constraints on galactic wind models}

\author[ A. Meiksin]{Avery Meiksin\\
SUPA\thanks{Scottish Universities Physics Alliance},
Institute for Astronomy, University of Edinburgh,
Blackford Hill, Edinburgh\ EH9\ 3HJ, UK}

\begin{document}

\maketitle

\begin{abstract}
  Observational implications are derived for two standard models of
  supernovae-driven galactic winds:\ a freely expanding steady-state
  wind and a wind sourced by a self-similarly expanding superbubble
  including thermal heat conduction. It is shown that, for the
  steady-state wind, matching the measured correlation between the
  soft x-ray luminosity and star formation rate of starburst galaxies
  is equivalent to producing a scaled wind mass-loading factor
  relative to the star-formation rate of 0.5--3, in agreement with the
  amount inferred from metal absorption line measurements. The match
  requires the asymptotic wind velocity $v_\infty$ to scale with the
  star formation rate $\dot M_*$ (in $M_\odot\,{\rm yr}^{-1}$)
  approximately as $v_\infty\simeq(700-1000)\,\kms\,{\dot
    M_*}^{1/6}$. The corresponding mass injection rate is close to the
  amount naturally provided by thermal evaporation from the wall of a
  superbubble in a galactic disc, suggesting thermal evaporation may
  be a major source of mass-loading. The predicted mass-loading
  factors from thermal evaporation within the galactic disc alone,
  however, are somewhat smaller, 0.2--2, so that a further
  contribution from cloud ablation or evaporation may be
  required. Both models may account for the 1.4~GHz luminosity of
  unresolved radio sources within starburst galaxies for plausible
  parameters describing the distribution of relativistic
  electrons. Further observational tests to distinguish the models are
  suggested.
\end{abstract}


\section{Introduction}
\label{sec:intro}

Galactic winds have been known to be common features of star-forming
galaxies for many years. While particularly spectacular winds, such as
those of M82, NGC~1482 and NGC~253 \citep[eg][]{1999ApJ...513..156M,
  2004ApJ...606..829S}, are exceptional, galaxies with modest winds
are widespread \citep{2005ARAA..43..769V}, as revealed by extraplanar
diffuse x-ray
\citep{1990ApJ...355..442F,1995ApJ...445..666A,2004ApJS..151..193S}
H$\alpha$ \citep{1990ApJS...74..833H,2003ApJS..148..383M}, and dust
emission \citep{1999AJ....117.2077H} in several galaxies.

Interest in the physical nature of the winds and their prevalence has
increased since it has been recognised they appear to play key roles
in galaxy formation and the distribution of metals in the
Intergalactic Medium (IGM). Both photoionization and mechanical
feedback from star-forming regions have long been expected to limit
star-formation on small scales \citep{2007ARAA..45..565M}. Wind
feedback has also been invoked to impede gas accretion and so limit
the efficiency of star formation in galaxy formation models to account
for disagreement between model predictions and observations
\citep[eg][]{1986ApJ...303...39D, 2012MNRAS.425.2027K,
  2015MNRAS.453.3499K, 2008MNRAS.387..577O,
  1998MNRAS.300..773W}. Winds extending over hundreds of kiloparsecs
may account for intergalactic metal absorption systems
\citep[see the review by][]{RevModPhys.81.1405}.

It is widely believed galactic winds result from the collective impact
of supernovae in compact star-forming regions on their surroundings
\citep[see the review by][]{2005ARAA..43..769V}. The details of the
physical mechanism driving the winds, however, are still unknown. Most
models are based on the injection of energy and matter by supernovae
and stellar winds in a distributed region \citep{1968MNRAS.140..241B,
  1971ApJ...165..381J, 1971ApJ...170..241M, 1985Natur.317...44C}. The
most straightforward model of pressure-driven expanding gas expelled
by the supernovae fails because it greatly over-predicts the gas
temperature as inferred from x-ray spectra of the winds, and
underpredicts the x-ray luminosities. Mass loading, in which
interstellar gas is incorporated into the flow, is recognized as the
most plausible explanation for the moderate temperatures and high
x-ray luminosities detected \citep{1993PASJ...45..513T,
  1996ApJ...463..528S}. This possibility receives observational
support from measurements of velocity-broadened metal absorption line
systems \citep{2015ApJ...809..147H}. Sources for the mass include
pre-existing interstellar gas in the vicinity of the supernovae from
stellar winds, hydrodynamic ablation of gas clouds entrained in the
outflow and thermal evaporation from gas clouds
\citep{1996ApJ...463..528S, 2000MNRAS.314..511S, 2001AA...367.1000P,
  2005MNRAS.362..626M}. Alternative wind models have also been
considered, such as momentum-driven winds \citep{2005ApJ...618..569M}
and winds driven by cosmic ray streaming \citep{1975ApJ...196..107I,
  2012MNRAS.423.2374U}.

Based on the stellar wind model of \citet{1975ApJ...200L.107C} and
\citet{1977ApJ...218..377W} for isolated stars, the galactic
superbubble model of \citet{1987ApJ...317..190M} and
\citet{1988ApJ...324..776M} naturally incorporates mass loading
sourced by thermal evaporation off the wall of the wind cavity
produced by supernovae, resulting in a self-similarly expanding
supershell. Analytic modelling and numerical computations suggest the
initially spherically expanding superbubble soon develops into a
biconal outflow within the stratified interstellar medium of a disc
galaxy \citep{1985ApJ...299...24S, 1988ApJ...324..776M,
  1988ApJ...330..695T}. Asymptotically, at large distances from the
galactic plane, the outflow may develop into a more spherical
superwind.

A more sophisticated approach to modelling winds is to use numerical
hydrodynamical computations to evolve a galactic wind from first
principles, but uncertainties in the detailed structure of the
interstellar medium limit the generality of the computations. The
computational demands imposed by the spatial resolution necessary to
capture all of the essential physics moreover precludes a fully
self-consistent treatment using cosmological simulations with current
resources. An element of \lq sub-grid physics' is ultimately required
\citep[eg][]{2003MNRAS.339..289S, 2012MNRAS.426..140D}. Critically,
all the numerical models suffer from one key deficiency:\ the unknown
physical mechanism driving the wind. In particular, the roles of both
clouds and thermal heat conduction, particularly as they affect the
amount of mass loading, are unknown. While the interaction between
winds and clouds has been investigated using numerical simulations
\citep[eg][]{2008ApJ...674..157C, 2015ApJ...805..158S}, most
simulations neglect thermal heat conduction \citep[but
  see][]{1999MNRAS.309..941D, 2005MNRAS.362..626M,
  2016arXiv160201843B}. The superbubble model has recently been
incorporated into cosmological simulations
\citep{2014MNRAS.442.3013K}, producing galaxy properties in good
agreement with observations \citep{2015MNRAS.453.3499K}.

A principal goal of this paper is to distinguish between models with
and without thermal heat conduction. The implicit justification for
neglecting heat conduction is that magnetic fields tangled by
turbulence will suppress heat conductivity. On the other hand, winds
will tend to comb out magnetic field lines, allowing some thermal heat
conduction along the wind direction. Which of these effects dominates
is not known.

The analytic modelling here is performed in the context of a
homogeneous wind. The homogeneity of wind gas is currently not well
constrained, although there is clear evidence for the presence of
entrained gas clouds. Even if initially the gas is clumped into
clouds, large mass-loading factors resulting from hydrodynamical
ablation or thermal evaporation may render the hot gas interior to the
wind bubble sufficiently smooth for the homogeneous models still to
provide a good approximation to the wind structure. Observations
suggest radiative losses, not included in the models, are generally
too small to affect the energetics of the winds, although they may
affect the metal column densities of embedded clouds
\citep{2001ApJ...554.1021H, 2002ApJ...577..691H}. Gravity will play a
secondary role in slowing the flow, an effect not readily incorporated
into an analytic treatment \citep[but see][]{2016ApJ...819...29B},
while it may limit the outflow velocity of ram-pressure driven clouds
\citep{2015ApJ...809..147H}. The principal role of gravity is in
stratifying the galactic disc gas, as will be discussed below.

The {\it Chandra X-ray Observatory} has enabled the development of
observing campaigns to systematically investigate the x-ray properties
of star-forming galaxies \citep{2004ApJS..151..193S,
  2005ApJ...628..187G, 2012MNRAS.419.2095M, 2012MNRAS.426.1870M,
  2014MNRAS.437.1698M}.
The primary quantity focussed on in this paper is the radiative x-ray
efficiency of the wind as quantified by the specific {\it diffuse}
x-ray energy generated per solar mass of stars formed. The x-ray
emission profiles extend to typical scales of several kiloparsecs
\citep{2004ApJS..151..193S, 2012MNRAS.426.1870M}, well beyond the
active star-forming regions driving the outflow. Winds on these scales
form bipolar cones through the stratified disc gas and become
increasingly inhomogeneous due to the onset of Kelvin-Helmholtz and
Rayleigh-Taylor instabilities, details not amenable to an analytic
treatment. Since the diffuse x-ray profiles are strongly centrally
peaked, however, the models considered here should still capture a
fair fraction of the total x-ray luminosity, particularly for dwarf
starbursts \citep{2005ApJ...628..187G}. This paper seeks to quantify
the radiative x-ray efficiency from the central region within the disc
of the galaxies, allowing for a range in star-formation rates and wind
properties. Although analytic models are approximate, they provide
insight into the origin of observational trends in terms of the
physical properties of the winds. They also provide invaluable
guidance into the design and interpretation of numerical simulations.

In the next section, approximate analytic scaling relations are
derived for the structure and some observational signatures of the
winds. In Sec.~\ref{sec:numeval_xray}, x-ray luminosity predictions
are presented using full numerical integrations of the models. This is
followed by predictions for radio luminosities in
Sec.~\ref{sec:numeval_rad} and for metal absorption column densities
in Sec.~\ref{sec:metals}. The results are discussed in
Sec.~\ref{sec:Discussion}, followed by a summary of the key results in
a conclusions section.

\section{Analytic estimates of specific x-ray emission}
\label{sec:estimates}

The x-ray emission from a wind arises from both thermal free-free and
line emission. To understand the dependence of the emission on the
properties of the sources and the surrounding gas, it is helpful first
to estimate the thermal free-free component analytically. In the
following section, more accurate results from numerical computations
are provided including line emission. Estimates use the model of
\citet{1985Natur.317...44C} for a steady-state wind and of
\citet{1987ApJ...317..190M} and \citet{1988ApJ...324..776M} for
superbubbles. Both assume gravitational acceleration is negligible, a
good approximation in the central regions of a galaxy where the gas is
much hotter than the galactic virial temperature. Gravity, however,
produces stratification of the galactic disc gas, which results in
biconal outflow. This is a limitation of both models:\ the
steady-state wind assumes a homogeneous source, limited therefore to a
region small compared with the scale-height of the disc. Comparison
with numerical simulations, however, suggest the wind produced by a
starburst in a cylindrical region in a disc may be rescaled to the
spherically symmetric solution to high accuracy
\citep{2009ApJ...697.2030S}. The superbubble model assumes a
homogeneous surrounding medium. Hydrodynamical simulations show this
tends to limit the growth of the superbubble to within the disc as
pressure-driven lobes emerge vertically
\citep{1989ApJ...337..141M}. The analysis here concentrates on the
structure of the superbubble when it reaches a size comparable to the
scale-height of the disc.

\subsection{Steady-state wind}
\label{subsec:ssw}

The characteristic ejecta energy and mass of a core-collapse supernova
are taken by \citet{1985Natur.317...44C} to be $E=10^{51}\,{\rm erg}$
and $M=3\,\msun$. Assuming a Salpeter stellar initial mass function
(IMF), a lower progenitor mass limit of $8\,\msun$, $\nu_{\rm SN}$ for
a core-collapse supernova gives a rate of about 1 core-collapse
supernova per 100 solar masses of stars formed. (The value would rise
by about 40 percent for a Kroupa IMF.)  Characterizing the supernova
rate as $0.01\nu_{100}$ supernova per solar mass of stars formed, the
energy and mass injection rates for a star-formation rate $\dot M_*$
are $\dot E=(10^{49}\epsilon\,{\rm erg}\, \msun^{-1})\nu_{100}\dot
M_*=\epsilon\dot E_1$ and $\dot M=(0.03\beta)\nu_{100}\dot
M_*=\beta\dot M_1$, where $\epsilon$ and $\beta$ allow for
uncertainties in the mechanical energy and mass-loading,
respectively. For $\dot M_*$ expressed in ${\rm M_\odot\, yr^{-1}}$,
$\dot E\simeq3.17\times10^{41}\,{\rm
  erg\,s^{-1}}\,\epsilon\nu_{100}{\dot M_*}$.

Most of the bolometric thermal free-free emission originates in the
central source region at $r<R$. The analytic estimate is based on
emission from this region. Using the results in the Appendix for a
$\gamma=5/3$ gas, the central hydrogen density is

\begin{equation}
  n_{{\rm H}0}\simeq0.00658\,{\rm cm}^{-3}\,\frac{\beta^{3/2}}{\epsilon^{1/2}}\nu_{100}\dot M_* R_{100}^{-2},
\label{eq:nH0}
\end{equation}

\noindent where $\dot M_*$ is the star formation rate in units of
$\msun\,{\rm yr}^{-1}$ and $R_{100}$ is the radius of the star
forming region in units of 100~pc. The central temperature is

\begin{equation}
  T_0 = \frac{2}{5}\frac{\bar m}{k_{\rm B}}\frac{\epsilon}{\beta}{\dot
          E_1}{\dot M_1}^{-1}
      \simeq4.76\times10^8\,{\rm K}\,\frac{\epsilon}{\beta}
      \simeq14.2v_\infty^2,
\label{eq:T0}
\end{equation}

\noindent where $\bar m$ is the mean mass per particle, and in the
last expression the temperature is characterized by the asymptotic
wind velocity at $r\gg R$,

\begin{equation}
v_\infty = 2^{1/2}\frac{\epsilon^{1/2}}{\beta^{1/2}}{\dot
                E_1^{1/2}}{\dot M_1}^{-1/2}
\simeq5790\,{\rm km\,s^{-1}}\,\frac{\epsilon^{1/2}}{\beta^{1/2}}.
\label{eq:vinf}
\end{equation}
In terms of $v_\infty$, the central hydrogen density is
$n_{{\rm H}0}\simeq1.28\,{\rm cm^{-3}}\,(v_\infty/1000\,{\rm
  km\,s^{-1}})^{-3}\epsilon\nu_{100}{\dot M}_* R_{100}^{-2}$.

Radiative cooling places a lower limit on the asymptotic wind
velocity. The energy injection rate must exceed the cooling rate
within the central region of the wind. The cooling rate is
$n_en_{\rm H}\Lambda_R(T)$ for electron and hydrogen number densities
$n_e$ and $n_{\rm H}$, respectively, where
$\Lambda_R(T)\simeq1.0\times10^{-22}\,{\rm erg\,cm^3\,s^{-1}}
T_6^{-0.7}\zeta_m$
with $T_6=T/10^6$~K and $\zeta_m$ the metallicty relative to solar
\citep{1988ApJ...324..776M}. Requiring
${\dot E}> n_en_{\rm H}\Lambda_R(T)(4\pi R^3/3)$ imposes the robust
restriction
\begin{equation} 
v_\infty > 550\,{\rm km\,s^{-1}}\left(\epsilon\nu_{100}\zeta_m{\dot M_*}
  R_{100}^{-1}\right)^{0.14},
\label{eq:vinfmin}
\end{equation}
corresponding to the limit on the mass-loading factor
$\beta<110\epsilon^{0.73}(\nu_{100}\zeta_m{\dot M_*}/
R_{100})^{-0.27}.$
In the literature, a more commonly defined mass-loading factor is the
ratio of mass outflow rate to star-formation rate. In terms of this
ratio, designated here by $\beta_*(=0.03\beta\nu_{100})$, the cooling
restriction imposes
\begin{equation}
  \beta_*=\left(\frac{1000\kms}{v_\infty}\right)^2\epsilon\nu_{100}<3.3\left(\epsilon\nu_{100}\right)^{0.73}\left({\dot
      M_*}\zeta_m R_{100}^{-1}\right)^{-0.27}.
\label{eq:betas}
\end{equation}
A similar restriction is derived by \citet{2014ApJ...784...93Z}. The
mass injection rate is then limited to ${\dot M} <
3.3(\epsilon\nu_{100}{\dot M_*})^{0.73}(R_{100}/
\zeta_m)^{0.27}\,\msun\,{\rm yr}^{-1}$.

As shown in the Appendix, the core density and temperature are nearly
uniform within $r< 0.98R$. Approximating the density and temperature
as constant within $r<R$, the total thermal free-free luminosity is
\begin{equation}
  L^{\rm
    ff}_\nu\simeq3\times10^{16}\,{\rm erg\,s^{-1}\,Hz^{-1}}\,  e^{-0.0244e_{\rm keV}(\beta/\epsilon)}
  \frac{\beta^{7/2}}{\epsilon^{3/2}}(\nu_{100}{\dot M}_*)^2R_{100}^{-1},
\label{eq:Lnuffssw}
\end{equation}
for x-rays of energy $e_{\rm keV}$ in keV. Integrating
Eq.~(\ref{eq:Lnuffssw}) over the energy band $(e_1-e_2)$~keV, the
x-ray energy produced per solar mass of stars formed is then
\begin{eqnarray}
\frac{E_{[e_1-e_2]\,{\rm keV}}}{M_*}&\simeq&8\times10^{42}\,{\rm  
  erg}\,{M_\odot}^{-1}\,\nu_{100}^2{\dot M}_*R_{100}^{-1}\nonumber\\
&&\times\frac{\beta^{5/2}}{\epsilon^{1/2}}\Biggl[\exp\left(-0.0244\frac{\beta}{\epsilon}e_{1,{\rm keV}}\right)\nonumber\\
&&-\exp\left(-0.0244\frac{\beta}{\epsilon}e_{2,{\rm keV}}\right)\Biggr].
\label{eq:ExMsssw}
\end{eqnarray}  
The model predicts a linear increase with the star formation
rate. Taking $\epsilon=1$ and a typical mass-loading factor of
$\beta\simeq100$, corresponding to the asymptotic wind velocity
$v_\infty\simeq600\,{\rm km\,s^{-1}}$, gives
\begin{equation}
\frac{E_{[0.5-2]\,{\rm keV}}}{M_*}\simeq2\times10^{47}\,{\rm  
  erg}\,{M_\odot}^{-1}\,\nu_{100}^2{\dot M}_*R_{100}^{-1}. 
\label{eq:ExMssswex}
\end{equation} 

\subsection{Superbubble with thermal heat conduction}
\label{subsec:ssbtc}

Allowing for an ambient interstellar medium and equilibration of the
temperature interior to the bubble cavity by thermal heat conduction,
\citet{1987ApJ...317..190M} and \citet{1988ApJ...324..776M} model the
superbubble as a self-similar expanding stellar wind.
Normalized by the typical mechanical luminosity of an OB association,
$L=10^{38}L_{38}\,{\rm erg\,s^{-1}}$, and for an ambient hydrogen
density outside the wind bubble $n_{{\rm H}, 0}$, the bubble radius
increases, assuming no radiative losses, like
\begin{equation}
R_{\rm B}\simeq 66\,{\rm pc}\left(\frac{L_{38}t_6^3}{n_{{\rm H},0}}\right)^{1/5},
\label{eq:SB_rad}
\end{equation}
where $t_6$ is the age of the bubble in units of $10^6$~yr. Adopting
the thermal conductivity coefficient
$\kappa(T)=6\times10^{-7}f_T\,{\rm erg\,s^{-1}\,cm^{-1}\,K^{-7/2}}$,
including a possible conductivity suppression factor $f_T$, the
interior bubble temperature and ionized hydrogen number density are
given in terms of the similarity variable $x=r/R_{\rm B}$ for radius
$r$ by
\begin{equation}
T \simeq (5.2\times10^6\,{\rm K})f_T^{-2/7}L_{38}^{8/35}n_{{\rm H},0}^{2/35}t_6^{-6/35}(1-x)^{2/5}, 
\label{eq:SB_T}
\end{equation}
and
\begin{equation}
n_{\rm H} \simeq (0.016\,{\rm cm^{-3}})f_T^{2/7}L_{38}^{6/35}n_{{\rm H}, 0}^{19/35}t_6^{-22/35}(1-x)^{-2/5}.
\label{eq:SB_nH}
\end{equation}

The bubble will cool primarily by line radiation at its surface. The
characteristic radiative cooling time is
\begin{equation}
t_{\rm R}\simeq(15\times10^6\,{\rm yr})L_{38}^{3/11}n_{{\rm
    H},0}^{-8/11}f_T^{-25/22}(\zeta_m+0.15)^{-35/22}.
\label{eq:tcool}
\end{equation}
The factor 0.15 has been added to $\zeta_m$ to account for hydrogen
and helium cooling, where care is taken near the surface to ensure
cooling is cut off below the recombination temperatures for helium and
hydrogen for collisionally ionized gas, and
$\Lambda_R(T)\sim T^{-1/2}$ was adopted for the surface layer,
following \citet{1988ApJ...324..776M}. For high ambient hydrogen
densities, cooling will limit the radius of the bubble to be smaller
than the characteristic scale height of the stratified interstellar
medium perpendicular to the disc. At lower densities, the bubble
radius may reach the disc scale height. The wind will then evolve into
a bipolar outflow perpendicular to the disc, and the expansion into
the plane of the disc ceases, or may even reverse
\citep{1988ApJ...324..776M, 1989ApJ...337..141M}.

Expressing the temperature and density of the gas interior to the
bubble as $T=T_cu^{2/5}$ and $n_{\rm H}=n_{{\rm H}c}u^{-2/5}$, where $u=1-x$,
the thermal free-free emission emitted by a wind bubble of radius
$R_{\rm B}=100R_{\rm B, 100}$~pc is
\begin{align}
L^{\rm ff}_\nu &\simeq& 5\times10^{25}\,{\rm 
  erg\,s^{-1}\,Hz^{-1}}n_{{\rm H}c}^2T_c^{-1/2}R_{\rm B, 100}^3\nonumber\\
&\times&\int_0^1\,\frac{du}{u}(1-u)^2\exp\left[-\left(\frac{h_{\rm P}\nu}{k_{\rm
   B}T_c}\right)u^{-2/5}\right] \label{eq:Lnuffssbtc}\\
&\simeq&\begin{cases}
10^{26}\,{\rm erg\,s^{-1}\,Hz^{-1}}n_{{\rm 
         H}c}^2T_c^{-1/2}R_{\rm B, 100}^3\nonumber\\
\times\left[\log\left(\frac{k_{\rm 
         B}T_c}{h_{\rm P}\nu}\right)-\frac{3+5\gamma}{5}\right]
         & (h_{\rm P}\nu\ll k_{\rm B}T_c);\nonumber\\
10^{27}\,{\rm erg\,s^{-1}\,Hz^{-1}}n_{{\rm 
         H}c}^2T_c^{-1/2}R_{\rm  B, 100}^3\nonumber\\
\times\left(\frac{k_{\rm 
         B}T_c}{h_{\rm P}\nu}\right)^3e^{-h_{\rm P}\nu/k_{\rm 
         B}T_c} &(h_{\rm P}\nu\gg k_{\rm B}T_c),
\end{cases}
\end{align}
where $h_{\rm P}$ is the Planck constant, $k_{\rm B}$ is the Boltzmann
constant, here $\gamma\simeq0.5772$ is Euler's constant, and the
integral has been evaluated with its asymptotic leading order
behaviour retained for the two limiting cases shown. A characteristic
central temperature of $\sim4\times10^7$~K gives a transition energy
between the two cases of about 3~keV.

At high densities, the growth of the wind bubble will be limited by
cooling once the energy radiated matches the total mechanical energy
deposited by the wind. This may be quantified as follows. At the
cooling time $t=t_{\rm R}$, the central hydrogen density and gas
temperature take on the values
\begin{equation}
n_{{\rm H}c}\simeq0.003\,{\rm cm^{-3}}\,n_{{\rm H},0}f_T(\zeta_m+0.15),
\label{eq:nHcc}
\end{equation}
independent of $L_{38}$, and
\begin{equation}
T_c\simeq1.4\times10^7\,{\rm K}\,(\epsilon\nu_{100} n_{{\rm H},0}\dot
M_*)^{2/11}f_T^{-1/11}(\zeta_m+0.15)^{3/11},
\label{eq:Tcc}
\end{equation}
respectively, where $L_{38}$ has been converted to the star formation
rate $\dot M_*$ (${\rm M_\odot\,yr^{-1}}$) using
$L_{38}\simeq3170\epsilon\nu_{100}{\dot M_*}$
(Sec.~\ref{subsec:ssw}). The cooling radius may be expressed as
\begin{equation} 
R_{\rm B, cool}\simeq6.3\,{\rm kpc}\,{n_{{\rm H},0}}^{-7/11}{\epsilon\nu_{100}\dot  
  M_*}^{4/11}f_T^{-15/22}(\zeta_m+0.15)^{-21/22}.  
\label{eq:Rbc}
\end{equation} 
The corresponding bubble expansion velocity
$\dot R_{\rm B, cool}=(3/5)R_{\rm B, cool}/t_{\rm R}$ at this time is
\begin{equation}
\dot R_{\rm B, cool}\simeq27\,{\rm km\,s^{-1}}\,(\epsilon\nu_{100}{\dot 
  M_* n_{{\rm H},0}})^{1/11}f_T^{5/11}(\zeta_m+0.15)^{7/11}. 
\label{eq:vbc}
\end{equation}
The mass interior to the bubble is dominated by the evaporation off
the bubble wall into the hot cavity at the rate
$\dot M_{\rm ev}=(16\pi/25)[{\bar m}\kappa(T_c)/ k_{\rm B}]R_{\rm B}$
\citep{1975ApJ...200L.107C}, where $\kappa(T_C)$ is the thermal
conductivity coefficient. The mass loading factor in the wind core
referenced to the star formation rate becomes
\begin{eqnarray}
\beta_*&=&\frac{\dot M_{\rm ev}}{\dot M_*}\simeq2.1\left({\dot M_*}n_{{\rm  
      H},0}\right)^{-2/11}\nonumber\\
&&\times (\epsilon\nu_{100})f_T^{1/11}(\zeta_m+0.15)^{-3/11}.  
\label{eq:betastcc}
\end{eqnarray}  
From Eq.~(\ref{eq:Lnuffssbtc}), the x-ray energy in the 0.5--2~keV  
band per solar mass of stars formed is then
\begin{equation}
\frac{E_{[0.5-2]\,{\rm keV}}}{M_*}\simeq2\times10^{47}\,{\rm  
  erg}\,{M_\odot}^{-1}\,(\epsilon\nu_{100})(\zeta_m+0.15)^{-1},
\label{eq:ExMssswtcc}
\end{equation} 
independent of the star formation rate, the ambient hydrogen density
and the rate of thermal heat conduction. It corresponds to 2 percent
of the mechanical energy radiated as x-rays in this band. The wind
will not immediately cease as the momentum of the outflow will
continue to sweep up material, but at a reduced rate
\citep{1992ApJ...388...93K}.

At lower densities, the bubble radius will be limited by the scale
height of the gas perpendicular to the plane. The central hydrogen
density and gas temperature when the bubble reaches a radius $R_{\rm B}$ are
\begin{equation}
n_{{\rm H}c}\simeq0.23\,{\rm cm}^{-3}\,(\epsilon\nu_{100})^{8/21}f_T^{2/7}n_{{\rm H},0}^{1/3}{\dot
  M_*}^{8/21}R_{\rm B, 100}^{-22/21}
\label{eq:nHw}
\end{equation} 
and
\begin{equation}
  T_c\simeq4.6\times10^7\,{\rm K}\,\left(\epsilon\nu_{100}{\dot M_*}/
    f_TR_{\rm B, 100}\right)^{2/7},
\label{eq:Tcw}
\end{equation} 
respectively. The corresponding wind velocity is
\begin{equation}
{\dot R}_{\rm B}\simeq430\,{\rm km\,s^{-1}}\,\left(\epsilon\nu_{100}{\dot 
  M_*}/n_{{\rm H},0}R_{\rm B, 100}^2\right)^{1/3}. 
\label{eq:vww}
\end{equation} 
The mass loading factor in the bubble is
\begin{equation}
\beta_*\simeq0.64(\epsilon\nu_{100})^{5/7}\left(R_{\rm B, 100}f_T/{\dot M_*}\right)^{2/7},
\label{eq:betastcsc}
\end{equation}  
independent of the ambient gas density. The x-ray energy in the
0.5--2~keV band per solar mass of stars formed is then
\begin{eqnarray}
\frac{E_{[0.5-2]\,{\rm keV}}}{M_*}&\simeq&5\times10^{45}\,{\rm  
  erg}\,{M_\odot}^{-1}\,(\epsilon\nu_{100})^{13/21}f_T^{5/7}\nonumber\\
&&\times n_{{\rm H},0}^{2/3}{\dot M_*}^{-8/21}R_{\rm B, 100}^{22/21},
\label{eq:ExMssswtc}
\end{eqnarray} 
decreasing weakly with increasing star formation rate.

The x-ray energy produced per solar mass of stars formed may then take
on a wide range of values, depending on $n_{{\rm H},0}$. For
$n_{{\rm H,0}}>n_{\rm H,0,R}\simeq670\,{\rm
  cm^{-3}}\,(\epsilon\nu_{100}\dot
M_*)^{4/7}f_T^{-15/14}(\zeta_m+0.15)^{-11/7}R_{\rm B, 100}$,
the rate will be near $10^{47}\,{\rm erg\, M_\odot^{-1}}$, where it
reaches a peak value independent of the ambient hydrogen density and
the star formation rate once radiative cooling restricts the bubble
growth.
 
It is instructive to compute the thermal heat conduction saturation
parameter for these two limiting cases. Following
\citet{1977ApJ...211..135C}, a consideration of the ratio of the mean
free path of the electrons to the temperature scale height for the
wind, expressed as an \lq inverted cloud,' shows that the surrounding
density and temperature in the cloud case should be replaced by the
central temperature and density of the wind. For a wind limited by
radiative cooling, the saturation parameter becomes
$\sigma_0=(T_c/1.54\times10^7\,{\rm K})^2f_T/(n_{\rm
  Hc}R_{\rm B}\phi_s)\simeq(0.05/\phi_s)f_T^{1/2}(\zeta_m+0.15)^{1/2}$,
where $\phi_s\sim1$ characterizes the uncertainty in the saturated
heat flux. This is nearly identical to the value
\citet{1977ApJ...215..213M} derive for interstellar clouds, below
which clouds will cool and condense rather than evaporate. If the wind
bubble is limited instead by the scale height of the disc to a radius
$100\,{\rm pc}\,R_{\rm B, 100}$, the saturation parameter becomes
$\sigma_0\simeq(0.39/\phi_s)f_T^{1/7}(\epsilon\nu_{100}\dot
M_*)^{4/21}n_{\rm H,0}^{-1/3}R_{\rm B, 100}^{-11/21}$.
Thermal heat conduction is thus close to being saturated
($\sigma_0>1$) for typical values of the parameters. Only models with
$\sigma_0<1$ are considered here so that the classical heat conduction
description applies.

In the following section, more precise numerical predictions are made
for the models, including the contribution from metal emission
lines. Comparisons with observations are also drawn.

\section{Numerical evaluation of specific x-ray emission}
\label{sec:numeval_xray}

\subsection{Data and modelling}
\label{subsec:model_xray}

The high angular and spectral resolution of the {\it Chandra X-ray
  Observatory} have enabled quantification of the correlation between
the soft x-ray diffuse emission associated with star forming regions
within galaxies and the star formation rate. From measurements of 6
disc galaxies, \citet{2009MNRAS.394.1741O} find
$L_{[0.3-1]\,{\rm keV}}\simeq (2-10)\times10^{38}\,{\rm
  erg\,s^{-1}}\dot M_*\,({\rm M_\odot\, yr^{-1}})$.
Only the most luminous point sources were removed, so that their value
may be conservatively viewed as an upper limit to the x-ray luminosity
of a gas component. \citet{2013MNRAS.435.3071L} find a similar
correlation between diffuse galactic coronal emission, corrected for
observed or estimated stellar contributions, and the star formation
rate of
$L_{[0.5-2]\,{\rm keV}}\simeq1.4^{+1.1}_{-0.8}\times10^{39}\,{\rm
  erg\,s^{-1}}\dot M_*\,({\rm M_\odot\, yr^{-1}})$
for 53 nearby disc galaxies. Based on star formation rate estimates
from infra-red and UV measurements restricted to the same projected
region as the diffuse x-ray emission, in a sample of galaxies cleaned
of those showing evidence of an active nucleus and with detected or
the estimated contribution of unresolved high mass x-ray binaries
removed, \citet{2012MNRAS.426.1870M} obtained
$L_{[0.5-2]\,{\rm keV}}\simeq(8.3\pm0.1)\times10^{38}\,{\rm
  erg\,s^{-1}}\dot M_*\,({\rm M_\odot\, yr^{-1}})$
for a sample of 21 late-type galaxies. (A Salpeter stellar initial
mass function was assumed.) On fitting a two-component thermal model
to the spectra, they find a correlation between the gaseous
contribution to the diffuse x-ray luminosity and the star formation
rate of
\begin{equation}
L_{[0.5-2]\,{\rm keV}}\simeq(5.2\pm0.2)\times10^{38}\,{\rm erg\,s^{-1}}\dot  
M_*\,({\rm M_\odot\, yr^{-1}}),
\label{eq:LxSFR}
\end{equation}
or  
$E_{[0.5-2]\,{\rm keV}}/M_*\simeq(1.6\pm0.1)\times10^{46}\,{\rm erg\,
  M_*^{-1}}$.  
For 9 galaxies, they find spectral evidence for substantial absorption
internal to the galaxies. Using these systems, they estimate the {\it
  intrinsic} diffuse gaseous emission to be
\begin{equation}
L_{[0.3-10]\,{\rm keV}}\simeq(7.3\pm1.3)\times10^{39}\,{\rm erg\,s^{-1}}\dot 
M_*\,({\rm M_\odot\, yr^{-1}}),
\label{eq:LxSFRcorr}
\end{equation}
or 
$E_{[0.3-10]\,{\rm keV}}/M_*\simeq(2.3\pm0.4)\times10^{47}\,{\rm erg\,
  M_*^{-1}}$. 

Since the x-ray emission in the wind models peaks within the energy
bands used to measure the emission, a more precise comparison between
the models and the measurements requires numerical integration of the
models. In addition to thermal free-free, x-ray line emission also
contributes substantially to the overall x-ray budget. The
\texttt{CHIANTI} rates \citep{2012ApJ...744...99L} for collisionally
ionized gas are adopted from Cloudy (13.03)
\citep{2013RMxAA..49..137F}, and emission tables for solar and
half-solar metallicity computed. Numerical integrations of the models
interpolate on the tables. Comparisons with measurements are made
separately below for the steady-state wind model and the superbubble
model.

\subsection{Steady-state wind}
\label{subsec:numssw_xray}

\begin{figure*}
\scalebox{0.45}{\includegraphics{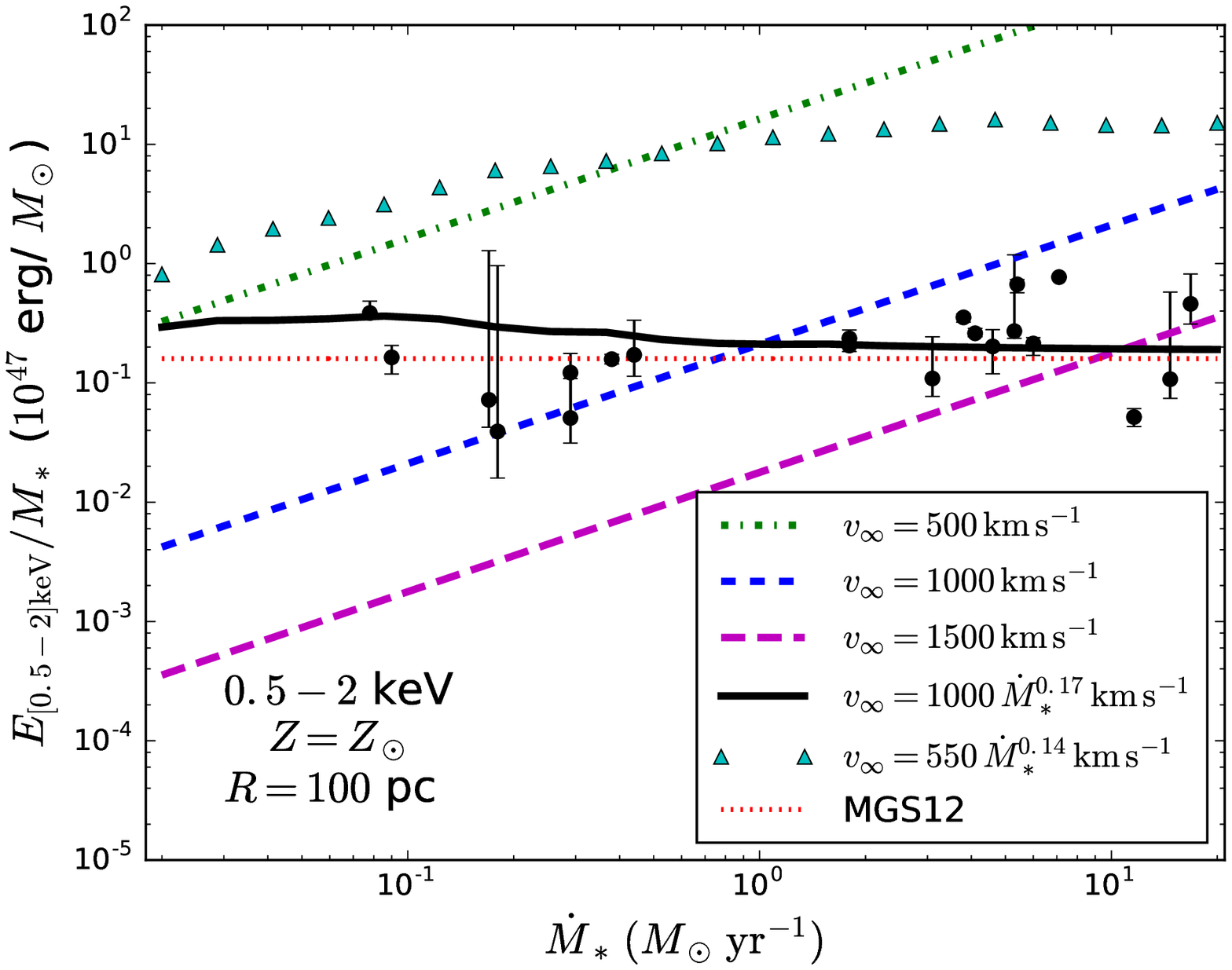}\includegraphics{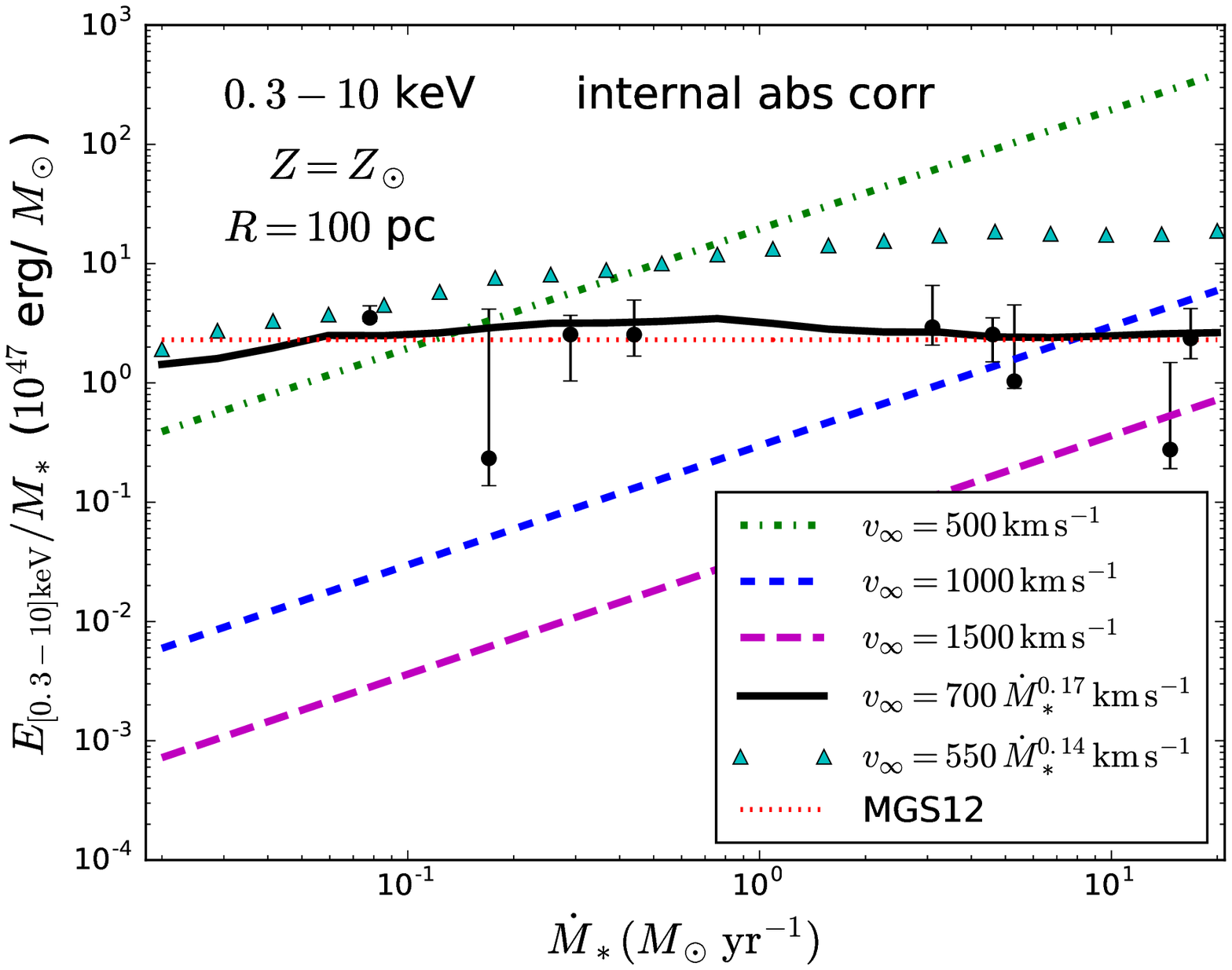}}
\caption{X-ray emission per solar mass of stars formed for a
  steady-state wind. Left panel:\ Shown for the x-ray band $0.5-2$~keV
  as a function of the star formation rate, for asymptotic wind
  velocities $v_\infty=500$, 1000 and 1500~${\rm km\,s^{-1}}$. The
  data points are from \citet{2012MNRAS.426.1870M}.  The error bars
  represent uncertainties in the distances to the galaxies. Right
  panel:\ As in the left panel, but for the x-ray band
  $0.3-10$~keV. The data points are from \citet{2012MNRAS.426.1870M},
  including correction for internal absorption. The triangles indicate
  the upper bound imposed by radiative cooling (see text). A source
  region 100~pc in radius and solar metallicity are assumed for both
  panels.
}
\label{fig:EmissSSW}
\end{figure*}

The specific x-ray emission in the bands $0.5-2$~keV and $0.3-10$~keV
for the steady-state wind model is shown in Fig.~\ref{fig:EmissSSW} as
a function of star formation rate. A source region of radius
$R=100$~pc is adopted, with solar metallicity. Emission from outside
the source region is included, although it diminishes rapidly with
distance outward. The mass-loading factor is expressed in terms of the
asymptotic wind velocity $v_\infty$.

The specific emissivity is a decreasing function of
$v_\infty$.  Expanding the source region to $R=200$~pc is found to
decrease the emission in the $0.5-2$~keV band by about 30 percent, the
same trend, but with a somewhat weaker dependence, as predicted by
Eq.~(\ref{eq:ExMsssw}). The x-ray emission is diminished by 30--50
percent on going from solar to half-solar metallicity for
$v_\infty>500\,{\rm km\,s^{-1}}$.

In the broader energy band $0.3-10$~keV, the specific emissivity,
shown in the right panel of Fig.~\ref{fig:EmissSSW}, decreases with
increasing volume of the source region, varying nearly as rapidly as
$1/R$, as in Eq.~(\ref{eq:ExMsssw}).  The specific emissivity
varies nearly linearly with metallicity, except for
$v_\infty\gsim1000\,{\rm km\,s^{-1}}$, for which the specific
emissivity depends only weakly on the metallicity.

The predicted linearly increasing trend with star formation rate is
not consistent with the observations. The data from
\citet{2012MNRAS.426.1870M} suggests a constant amount of x-ray energy
emitted per unit mass of star formed. In the energy band $0.5-2$~keV,
this is matched by allowing a tight correlation between the asymptotic
wind velocity and the star formation rate according to
$v_\infty\simeq1000 {\dot M_*}^{1/6}\,\kms$, corresponding to a
central hydrogen density
$n_{\rm H,0}\simeq1.3 {\dot M_*}^{1/2}\,{\rm cm^{-3}}$. This results
in an increasing amount of mass loading for a decreasing star
formation rate, a general requirement recognized by
\citet{2014ApJ...784...93Z}.

Results for galaxies corrected for internal absorption are shown in
the right hand panel of Fig.~\ref{fig:EmissSSW} for the band
$0.3-10$~keV. The near constancy of the specific emissivity with star
formation rate persists in the data. Agreement with the data may again
be achieved if the wind velocity were tightly correlated with the star
formation rate according to
$v_\infty\simeq700{\rm km\,s^{-1}}{\dot M_*}^{1/6}$, corresponding to
a central hydrogen density
$n_{\rm H,0}\simeq4 {\dot M_*}^{1/2}\,{\rm cm^{-3}}$. The scaling with
star formation rate is bracketed by that expected from
Eqs.~(\ref{eq:Lnuffssw}) and (\ref{eq:ExMsssw}), which give for a
constant luminosity per rate of star formation the approximate
analytic scaling
$v_\infty\sim(\epsilon^2\nu_{100}^2{\dot M_*}/R_{100})^\alpha$ with
$\alpha=1/7-1/5$. The velocity correlations are close to the cooling
restriction Eq.~(\ref{eq:vinfmin}), suggesting a narrow range is
allowed for viable winds \citep[cf][]{2014ApJ...784...93Z}.

\subsection{Superbubble with thermal heat conduction}
\label{subsec:numsbtc_xray}

\begin{figure*}
\scalebox{0.45}{\includegraphics{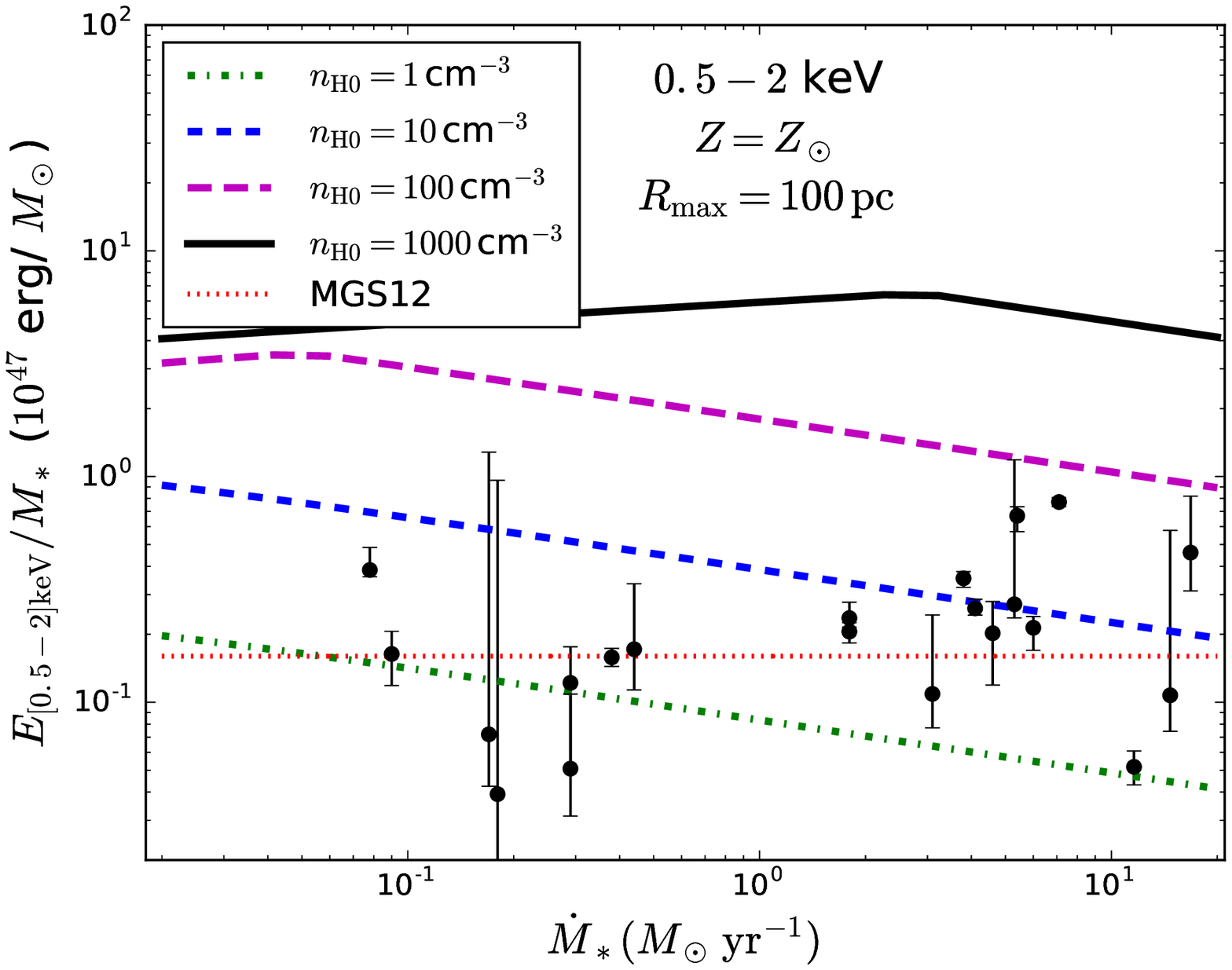}\includegraphics{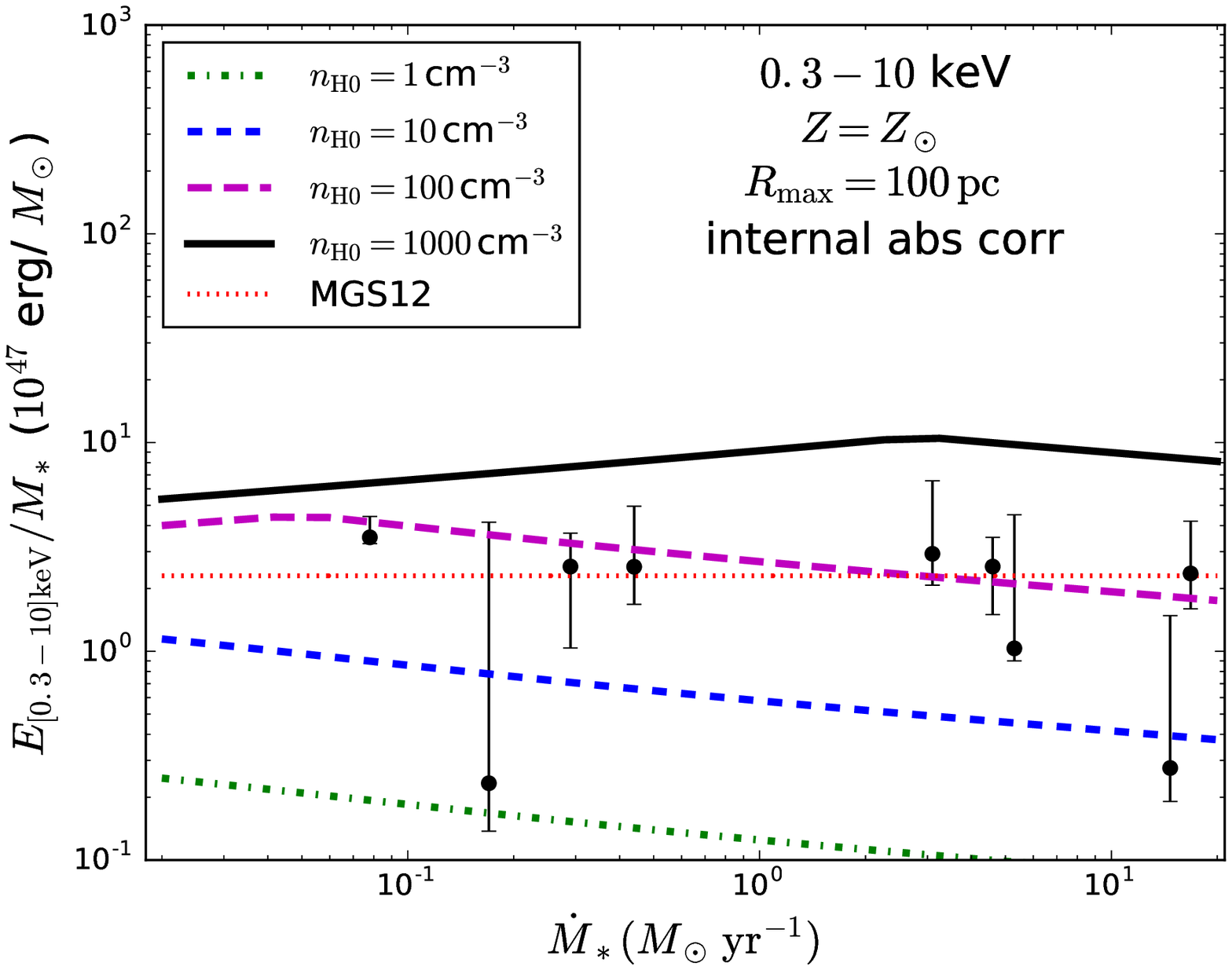}}
\caption{X-ray emission per solar mass of stars formed for a
  superbubble including thermal heat conduction. Left panel:\ Shown
  for the x-ray band $0.5-2$~keV as a function of the star formation
  rate, for external hydrogen densities $n_{\rm H,0}=1$, 10, 100 and
  1000~${\rm cm\,s^{-3}}$. The data points are from
  \citet{2012MNRAS.426.1870M}. Right panel:\ As in the left panel, but
  for the x-ray band $0.3-10$~keV. The data points are from
  \citet{2012MNRAS.426.1870M}, including correction for internal
  absorption.  The error bars represent uncertainties in the distances
  to the galaxies. A maximum wind radius of 100~pc and solar
  metallicity are assumed for both panels.
}
\label{fig:EmissSSTC}
\end{figure*}

Allowing for thermal evaporation from a surrounding medium results in
a much narrower range in specific emissivity compared with the steady
state model. Results assuming a maximum radius of 100~pc for the
expanding bubble and for solar metallicity are shown in
Fig.~\ref{fig:EmissSSTC}. The results shown are time-averaged over the
duration of the spherical expansion of the wind, assumed to cease at
the cooling time, Eq.~(\ref{eq:tcool}), or when it reaches the maximum
radius. X-ray emission only from within the maximum radius is
computed. The emission will fall off rapidly away from the plane if
the superbubble expands out of the disc, but emission from an extended
region may be comparable to that from within the disc. A
multi-dimensional model is required to estimate the full emission more
accurately.

The specific x-ray emissivity increases with the ambient hydrogen
density approximately as $n_{\rm H,0}^{0.7}$, in agreement with
Eq.~(\ref{eq:ExMssswtc}), except at the highest density and low star
formation rate where the wind expansion is cooling limited. At high
densities, the specific emissivity becomes nearly independent of the
star formation rate and gas density, in agreement with
Eq.~(\ref{eq:ExMssswtcc}).

For low star formation rates and low ambient hydrogen densities, the
specific emissivity nearly halves on going from solar to half-solar
metallicity. The x-ray emission is dominated by line emission. The
difference is much more moderate at high star formation rates and high
ambient densities, for which line emission no longer dominates. At low
ambient hydrogen densities, the specific emissivity increases linearly
with the maximum bubble radius, but less rapidly at higher densities
as cooling becomes important, especially for low star formation rates,
in accordance with Eqs.~(\ref{eq:ExMssswtcc}) and
(\ref{eq:ExMssswtc}).

As shown in the left hand panel of Fig.~\ref{fig:EmissSSTC},
comparison with the measured specific emissivities using the data from
\citet{2012MNRAS.426.1870M}, assuming no internal absorption from the
galaxies, shows good agreement for ambient gas densities of
$n_{\rm H,0}\simeq1-10\,{\rm cm^{-3}}$. For a fixed star formation
rate, the required $n_{\rm H,0}$ will scale like $R_{\rm B}^{-11/7}$
according to Eq.~(\ref{eq:ExMssswtc}). Since the measured values
likely exceed the emission from the inner region within the disc by a
factor of a few, the implied hydrogen densities are likely somewhat
smaller. Allowing for internal absorption, agreement with the data in
the right hand panel shows values of
$n_{\rm H,0}\simeq10-100\,{\rm cm^{-3}}$ are preferred. No other
parameters need be adjusted:\ the model predicts the specific x-ray
emissivity is only weakly dependent on the star formation rate, in
agreement with the data.

\section{Specific radio emission}
\label{sec:numeval_rad}

\subsection{Data and modelling}
\label{subsec:model_rad}

The radio continuum radiation emitted by star-forming galaxies scales
with the star formation rate, at least for large radio luminosities
\citep{1992ARAA..30..575C, 2003ApJ...586..794B}. The physical origin
of the emission is unknown, but it is suspected to arise both from
shocks driven by stellar winds and supernovae and from cosmic rays in
a large-scale magnetic field. Measurements suggest that 90 percent of
the continuum emission at 1.4~GHz is synchrotron and 10 percent
thermal free-free in nature, suggesting a component from \HII\ regions
as well \citep{1992ARAA..30..575C}. Modelling all these effects is
well beyond the scope of this paper. Here only the synchrotron and
free-free radio emission from the wind regions are estimated. In
comparing with radio data, it is unclear from which scale to take the
emission. The correlation between the radio continuum and the star
formation rate is based on extended regions that likely include
emission from large-scale interstellar cosmic rays. A representative
value is
\begin{equation}
L_{1.4\,{\rm GHz}} ({\rm erg\,s^{-1}\,Hz^{-1}}) \simeq 8.4\times10^{27}\dot 
M_*\,({\rm M_\odot\, yr^{-1}})
\label{eq:LrSFR}
\end{equation}
\citep{2002AJ....124..675C}. By contrast, the dominant emission from
shocks within the wind region would be much more centrally
concentrated.

The {\it FIRST} radio survey \citep{1995ApJ...450..559B} includes data
that matches the scale of the x-ray and star-forming regions,
typically up to a few arcminutes, measured by
\citet{2012MNRAS.426.1870M}. We compare the models with two 1.4~MHz
continuum measurements from the {\it FIRST} survey, the large-scale
value centred on each galaxy and the brightest unresolved peak value
in the nucleus of the galaxy, corresponding typically to a region
within the central 100--500~pc of the galaxy for a source at a
distance of 10--20~Mpc. As shown below, the large scale values agree
well with Eq.~(\ref{eq:LrSFR}), corresponding to
$L_{1.4\,{\rm GHz}}/{\dot M_*}\simeq2.6\times10^{35}\,{\rm
  erg\,Hz^{-1}\,M_\odot^{-1}}$.

The synchrotron and thermal free-free emission are computed for the
models as in \citet{2013MNRAS.430.2854M}. In brief, a power-law energy
distribution $dn/d\epsilon\sim\epsilon^{-p_e}$ is assumed for the
relativistic electrons, with an energy density a fraction $f_e$ of the
local thermal energy density. The magnetic field energy density is
also taken to be $f_e$ for simplicity, corresponding to approximate
equipartition. The relativistic electrons are allowed to cool by
synchrotron and thermal free-free radiation and by excitation of
plasmon waves following the passage of the wind-driven shock front
into the interstellar gas. Thermal free-free and synchrotron
self-absorption are included, although for the frequencies of interest
these are generally negligible in the models considered. Observations
of supernova remnant spectra suggest typical values for the
relativistic electron energy index of $2<p_e<3$
\citep{1998ApJ...499..810C, 1986ApJ...301..790W}, while representative
model values for the relativistic electron energy density fraction
range over $0.001 < f_e < 0.2$ \citep{2006ApJ...641.1029C}. The
predictions of the wind models for radio emission are estimated to
check they do not exceed the observed limits for plausible
parameters. Virtually all the emission predicted by the models is
synchrotron radiation; the thermal free-free component is two to three
orders of magnitude smaller.

\subsection{Steady-state wind}
\label{subsec:numssw_rad}

\begin{figure*}
\scalebox{0.45}{\includegraphics{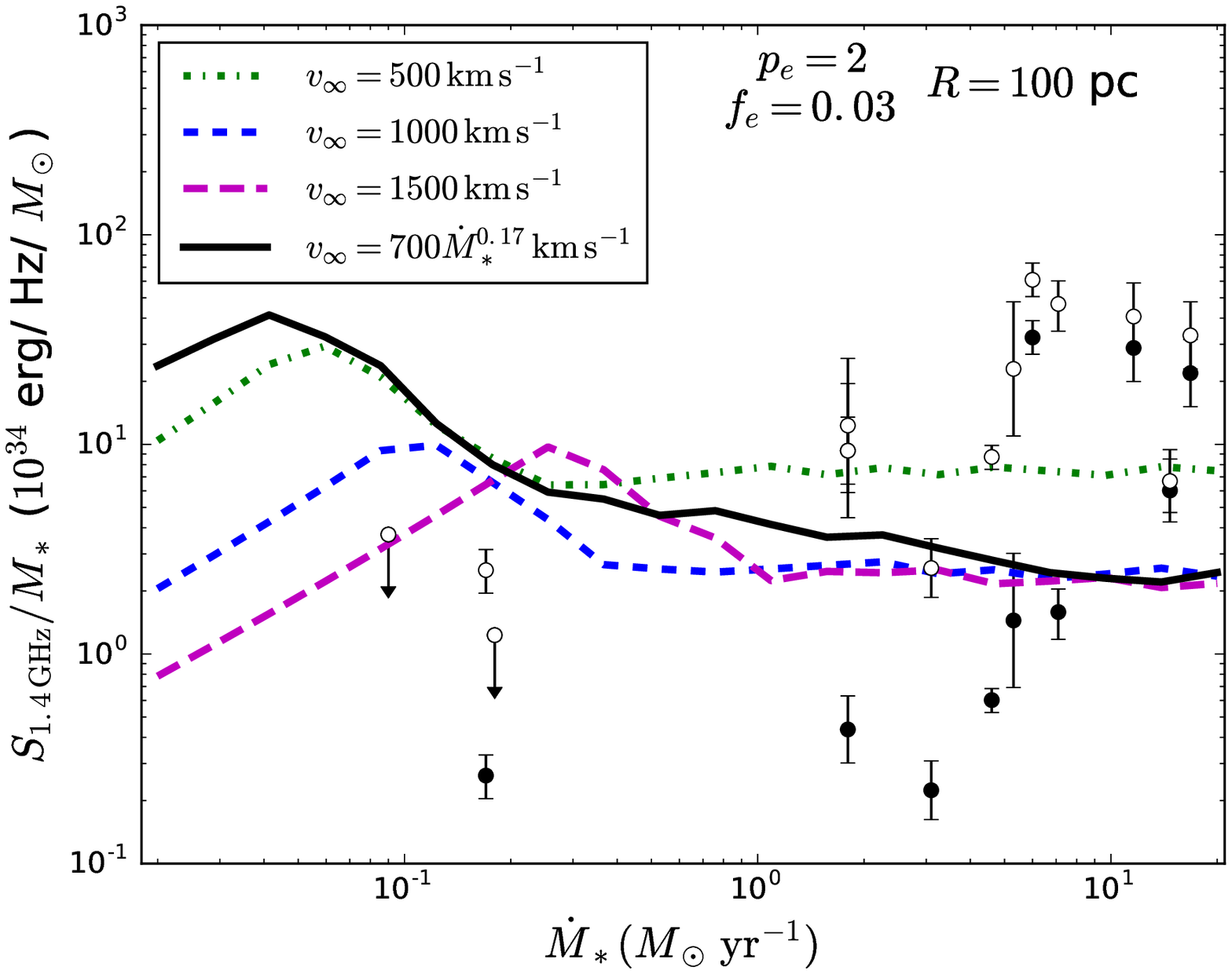}\includegraphics{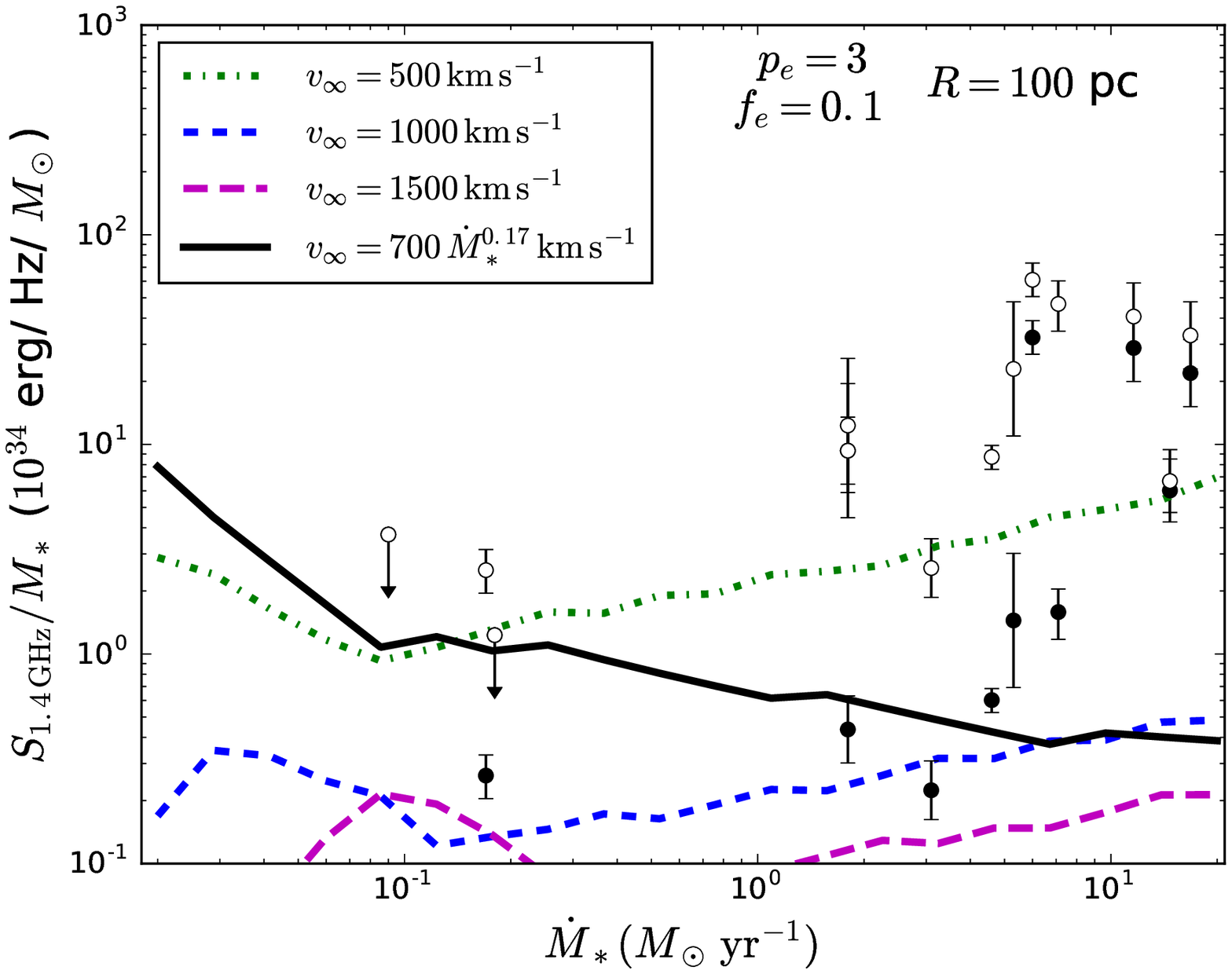}}
\caption{Radio continuum emission at 1.4~GHz per solar mass of stars
  formed for a steady-state wind. Shown for a relativistic electron
  energy index $p_e=2$ and energy density fraction $f_e=0.03$ (left
  panel) and $p_e=3$, $f_e=0.1$ (right panel), as a function of the
  star formation rate, for asymptotic wind velocities $v_\infty=500$,
  1000 and 1500~${\rm km\,s^{-1}}$. The emission adopting
  $v_\infty=700{\dot M_*}^{1/6}\,\kms$ is shown by the solid (black)
  lines. The data points are from the {\it FIRST} radio survey. The
  error bars represent distance uncertainties. Open points represent
  the large-scale emission; filled points represent the peak
  unresolved emission (see text). A wind source region 100~pc in
  radius is assumed for both panels.
}
\label{fig:RadioSSW}
\end{figure*}

The x-ray measurements require models with mass-loading from a gas
reservoir surrounding the supernovae. The wind will then drive a shock
into the surroundings. Since the wind is in a steady state, however,
the time since the shock passed a given radius is undetermined in the
model. To allow an estimate of the synchrotron emission, the wind is
arbitrarily assumed to have reached a distance of 10~kpc from the
source region, corresponding to an age of $t_{\rm age}=10\,{\rm
  kpc}/v_\infty$.

For a characteristic wind age of $10^7t_{w,7}\,{\rm yr}$, for low star
formation rates the synchrotron emission is dominated by the source
region $r<R$, giving, for $p_e=2$,
\begin{eqnarray}
\frac{S_{\rm 1.4GHz}}{\dot M_*}&\simeq&5\times10^{37}\,{\rm  
  erg\,Hz^{-1}\,M_\odot^{-1}}\,\nu_{\rm GHz}^{-1/2}f_e^{7/4}R_{100}^{-1/2}\nonumber \\
&&\times \left(\frac{1000\,{\rm  
      km\,s^{-1}}}{v_\infty}\right)^{7/4}\left(\epsilon\nu_{100}\right)^{7/4}
{\dot M_*}^{3/4}.
\label{eq:SnuMss}
\end{eqnarray} 
For a typical value $f_e=0.01$, $S_{\rm 1.4GHz}/{\dot
  M_*}\simeq1.2\times10^{34}\,{\rm erg\,Hz^{-1}\,M_\odot^{-1}}$.

At higher star formation rates, the density becomes sufficiently high
that two additional effects become important:\ synchrotron and plasmon
generation losses deplete the central region of relativistic
electrons; for sufficiently strong synchrotron losses the critical
frequency falls to several gigahertz or less. Emission then arises
only from outside the source region. The criterion for plasmon and
synchrotron losses not to deplete the population of relativistic
electrons is $f_en_{\rm H}Pt_{w,7}^2<8\times10^{-11}\,{\rm
  dyne\,cm^{-5}}$, where $P$ is the gas pressure in the wind. Using
the asymptotic limits for density and pressure, this requires emission
to arise only from radii
\begin{equation}
\frac{r}{R}>1.1f_e^{3/16}R_{100}^{-3/4}t_{w,7}^{3/8}\left(\epsilon\nu_{100}{\dot
    M_*}\right)^{3/8}\left(\frac{1000\,{\rm
      km\,s^{-1}}}{v_\infty}\right)^{3/4},
\label{eq:xspl}
\end{equation}
resulting in a rate of radio energy generation at frequency $10^9\nu_{\rm
  GHz}$~Hz per solar mass of stars formed of
\begin{eqnarray}
\frac{S_\nu}{\dot M_*}&\simeq&4\times10^{37}\,{\rm 
  erg\,Hz^{-1}\,M_\odot^{-1}}\, \nu_{\rm
                               GHz}^{-1/2}f_e^{39/32}R_{100}^{13/8}\nonumber\\
&&\times \left(\frac{v_\infty}{1000\,{\rm 
      km\,s^{-1}}}\right)^{3/8}\left(\epsilon\nu_{100}\right)^{11/16}{\dot  
  M_*}^{-5/16}t_{w,7}^{-17/16}. 
\label{eq:Snuspl}
\end{eqnarray} 
At even higher densities, once synchrotron losses lower the synchrotron
critical frequency into the gigahertz range, emission above
frequency $\nu_{\rm GHz}$ occurs only at radii
\begin{equation}
\frac{r}{R}>3.2\nu_{\rm GHz}^{1/5}f_e^{3/10}R_{100}^{1/5}t_{w,7}^{2/5}\left(\epsilon\nu_{100}{\dot
    M_*}\right)^{3/10}\left(\frac{1000\,{\rm
      km\,s^{-1}}}{v_\infty}\right)^{3/10},
\label{eq:xs}
\end{equation}
resulting in a rate of radio energy generation per solar mass of stars formed
\begin{eqnarray}
\frac{S_\nu}{\dot M_*}&\simeq&3\times10^{35}\,{\rm 
  erg\,Hz^{-1}\,M_\odot^{-1}}\, \nu_{\rm
                               GHz}^{-16/15}f_e^{9/10}R_{100}^{-16/15}\nonumber\\
&&\times \left(\frac{1000\,{\rm 
      km\,s^{-1}}}{v_\infty}\right)^{9/10}\left(\epsilon\nu_{100}\right)^{9/10}{\dot
  M_*}^{-1/10}t_{w,7}^{-17/15}. 
\label{eq:Snus}
\end{eqnarray}

Results from numerically integrating the model are shown in
Fig.~\ref{fig:RadioSSW}. The 1.4~GHz emission is displayed as a
function of the star-formation rate and $v_\infty$ for $p_e=2$,
$f_e=0.03$ and $p_e=3$, $f_e=0.1$. All three trends with the star
formation rate are apparent:\ a rise ($\dot M_*^{3/4}$), followed by a
decline ($\dot M_*^{-5/16}$) once synchrotron and plasmon losses pinch
off the relativistic electron distribution within the wind source
region, and finally a near constant level ($\dot M_*^{-1/10}$) once
synchrotron losses restrict the generation of 1.4~GHz power to regions
well outside the core. (The power at large star formation rates for
the $v_\infty=1500\kms$ case differs from the trend with $v_\infty$ in
Eq.~(\ref{eq:Snus}) because the radius at which the critical frequency
exceeds 1.4~GHz lies just outside the source region, where the
asymptotic decrease of pressure with radius is no longer a good
approximation.) For a high star formation rate, the radio spectrum for
$p_e=2$ is found to steepen to $\nu^{-\alpha_S}$ with
$\alpha_S\simeq1.5$, rather than the expected
$\alpha_S=(p_e-1)/2=1/2$, as is found for low star formation rates:\
the truncation of the emitting volume at high densities steepens the
spectrum.

The predicted trend adopting the correlation
$v_\infty\simeq700\kms\, {\dot M_*}^{1/6}$ is shown by the solid
(black) curves. For $p_e=2$, the fluxes for the unresolved radio
sources are largely recovered or exceeded. The high flux values for
the highest star-formation rates are matched only for
$f_e\simeq0.1-0.2$, the limiting values inferred for supernova
remnants. These high fluxes, however, are derived from distant
galaxies within regions unresolved on the scales of 1--2~kpc, so the
radio emission may be contaminated by emission from cosmic rays
interacting with large-scale galactic magnetic fields. The predicted
excess emission for unresolved sources with low flux values may
indicate a reduced volume filling factor of emitting electrons,
resulting in small volume-averaged values for the relativistic energy
fraction. At high star formation rates, the radio flux also decreases
inversely with the bubble size, according to
Eq.~(\ref{eq:Snus}). Alternatively, the relativistic electron energy
distribution may be steeper. For $p_e=3$, $f_e\approx0.1$ recovers the
lower flux values of the unresolved sources.

\subsection{Superbubble with thermal heat conduction}
\label{subsec:numsbtc_rad}

\begin{figure*}
\scalebox{0.45}{\includegraphics{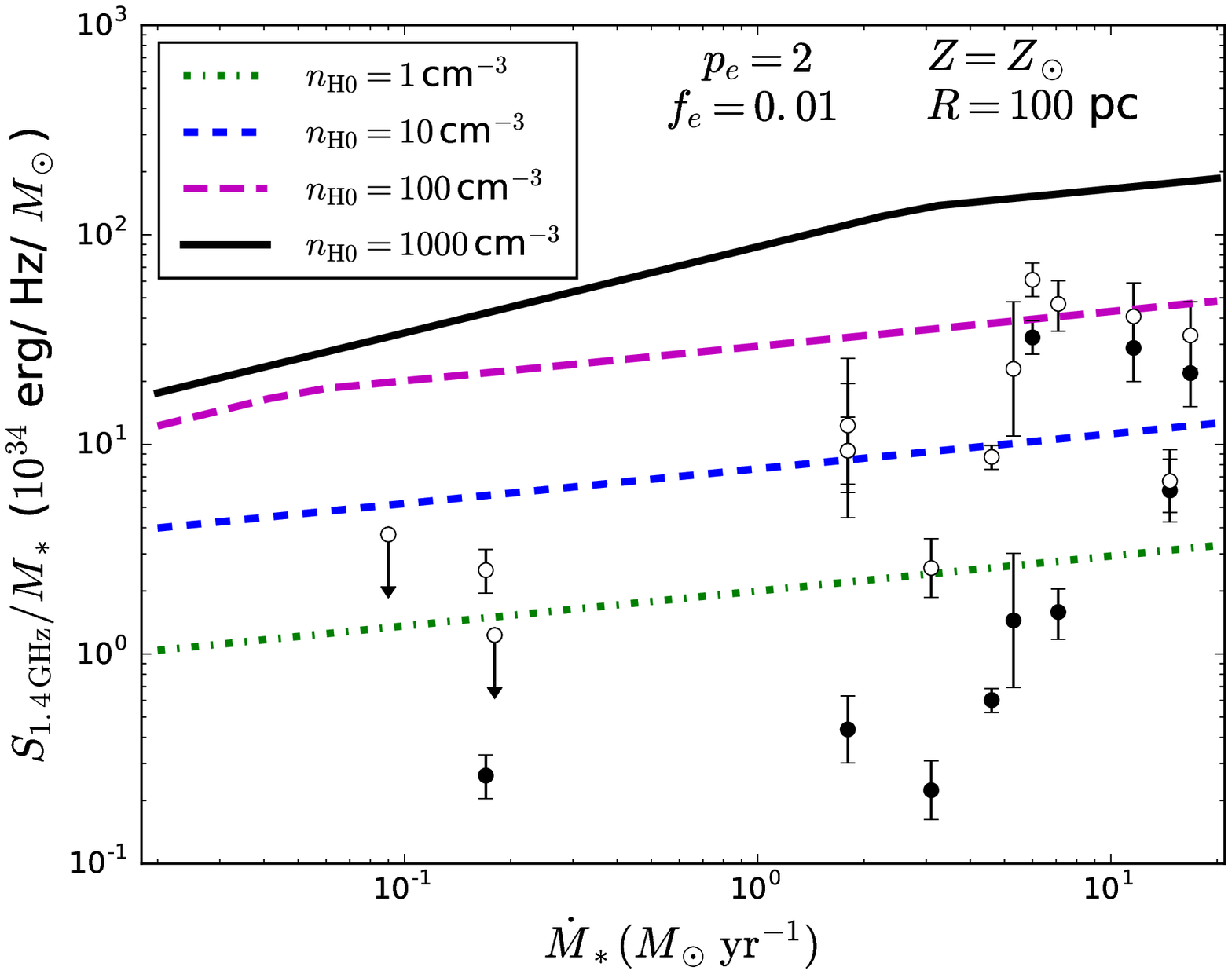}\includegraphics{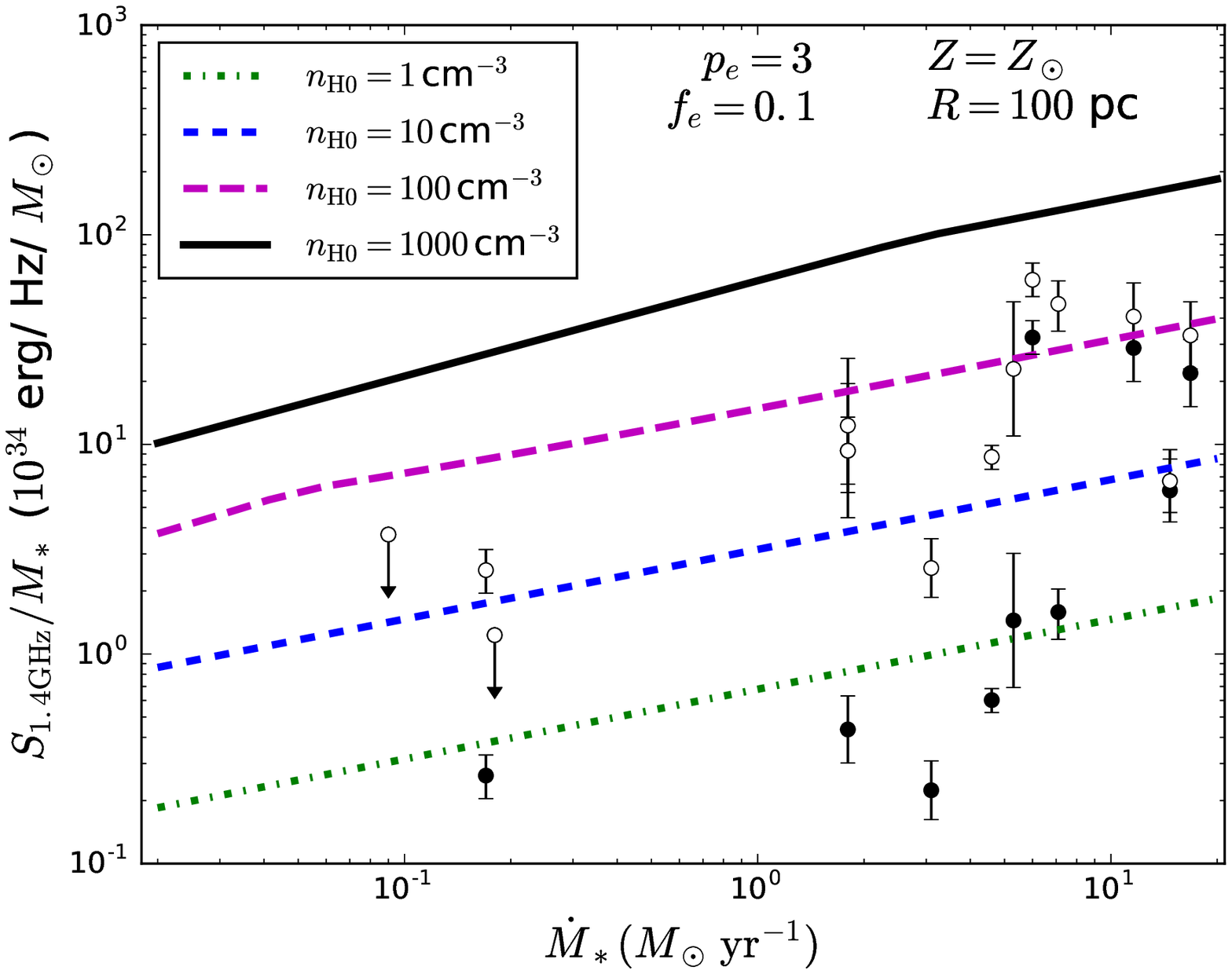}}
\caption{Radio continuum emission at 1.4~GHz per solar mass of stars
  formed for the superbubble model. Shown for relativistic electron
  distribution index $p_e=2$ and energy density fraction $f_e=0.01$
  (left panel) and $p_e=3$, $f_e=0.1$ (right panel), as a function of
  the star formation rate, for external hydrogen densities
  $n_{\rm H,0}=1$, 10, 100 and 1000~${\rm cm\,s^{-3}}$. The data
  points are from the {\it FIRST} radio survey. Open points represent
  the large-scale emission; filled points represent the peak
  unresolved emission (see text). A maximum bubble radius of 100~pc is
  adopted and solar metallicity.
}
\label{fig:RadioSSTC}
\end{figure*}

For the superbubble model, the characteristic radio emission for a
scale-height limited superbubble from within $R_{\rm B, 100}$ is given
approximately by
\begin{eqnarray}
\frac{S_\nu}{\dot M_*}&\simeq&1.0\times10^{38}\,{\rm 
  erg\,Hz^{-1}\,M_\odot^{-1}}\, \nu_{\rm
  GHz}^{-1/2}f_e^{7/4}\nonumber\\
&&\times(\epsilon\nu_{100})^{7/6}n_{\rm H,0}^{7/12}{\dot
  M_*}^{1/6}R_{\rm B, 100}^{2/3},
\label{eq:Snustc}
\end{eqnarray} 
nearly independent of the star formation rate. The radio emission is
independent of any suppression of thermal conductivity since the radio
power is determined by the thermal energy density, not the gas density
or temperature separately in the absence of significant
attenuation. Results from numerical integration of the wind equations
are shown in Fig.~\ref{fig:RadioSSTC}. The bubble region recovers the
full range of measured large-scale radio power for $p_e=2$ and
$f_e=0.01$. The larger flux values, at high star formation rates,
require a high ambient hydrogen density of
$n_{{\rm H},0}\sim100\,{\rm cm^{-3}}$, the upper value required by the
x-ray data. The lower power from unresolved regions is over-predicted
for the higher star formation rates, possibly indicating a reduced
volume filling factor of emitting electrons. The steeper electron
distribution case with $p_e=3$ requires increasing the relativistic
electron energy fraction to $f_e=0.1$, approaching the limit from
supernova remnant modelling. A near constant specific radio power is
found for a given ambient hydrogen density, only weakly dependent on
the star formation rate, in agreement with Eq.~(\ref{eq:Snustc}).

\section{Metal column densities}
\label{sec:metals}

\subsection{Data and modelling}
\label{subsec:model_metals}

An estimate of the column densities of metal ions within the winds may
be made as follows. For solar metallicity, abundances by number of
commonly detected metal atoms compared with hydrogen include:\
$\log_{10}\xi_{\rm He} = -1.07$, $\log_{10}\xi_{\rm C} = -3.57$,
$\log_{10}\xi_{\rm N} = -4.17$, $\log_{10}\xi_{\rm O} = -3.31$,
$\log_{10}\xi_{\rm Si} = -4.49$ and $\log_{10}\xi_{\rm S} = -4.88$
\citep{2009ARAA..47..481A}. Any given ionization state will dominate
at a particular temperature where it contributes most to the column
density of that ion, although for some species neighbouring ionization
states share substantially in the ionization. Temperatures at which
commonly measured ions peak include: \CIII at $10^{4.8}$~K
($10^{-3.6}$), \CIV at $10^{5.0}$~K ($10^{-4.1}$), \NII at
$10^{4.4}$~K ($10^{-4.2}$), \NIII at $10^{4.9}$~K ($10^{-4.2}$), \OVI
at $10^{5.5}$~K ($10^{-4.2}$), \SiIII at $10^{4.5}$~K ($10^{-4.6}$),
\SiIV at $10^{4.8}$~K ($10^{-5.0}$), SIII at $10^{4.7}$~K
($10^{-5.0}$) and \SIV at $10^{5.0}$~K ($10^{-5.1}$), where the peak
abundance fractions by number relative to hydrogen,
$\xi_{i,{\rm max}}$, are indicated in parentheses. It is noted these
values will be modified if the gas is not in collisional ionization
equilibrium, as appears to be the case for some high velocity clouds
in the Galactic halo \citep{2011ApJ...739..105S}.

\begin{figure*}
\scalebox{0.45}{\includegraphics{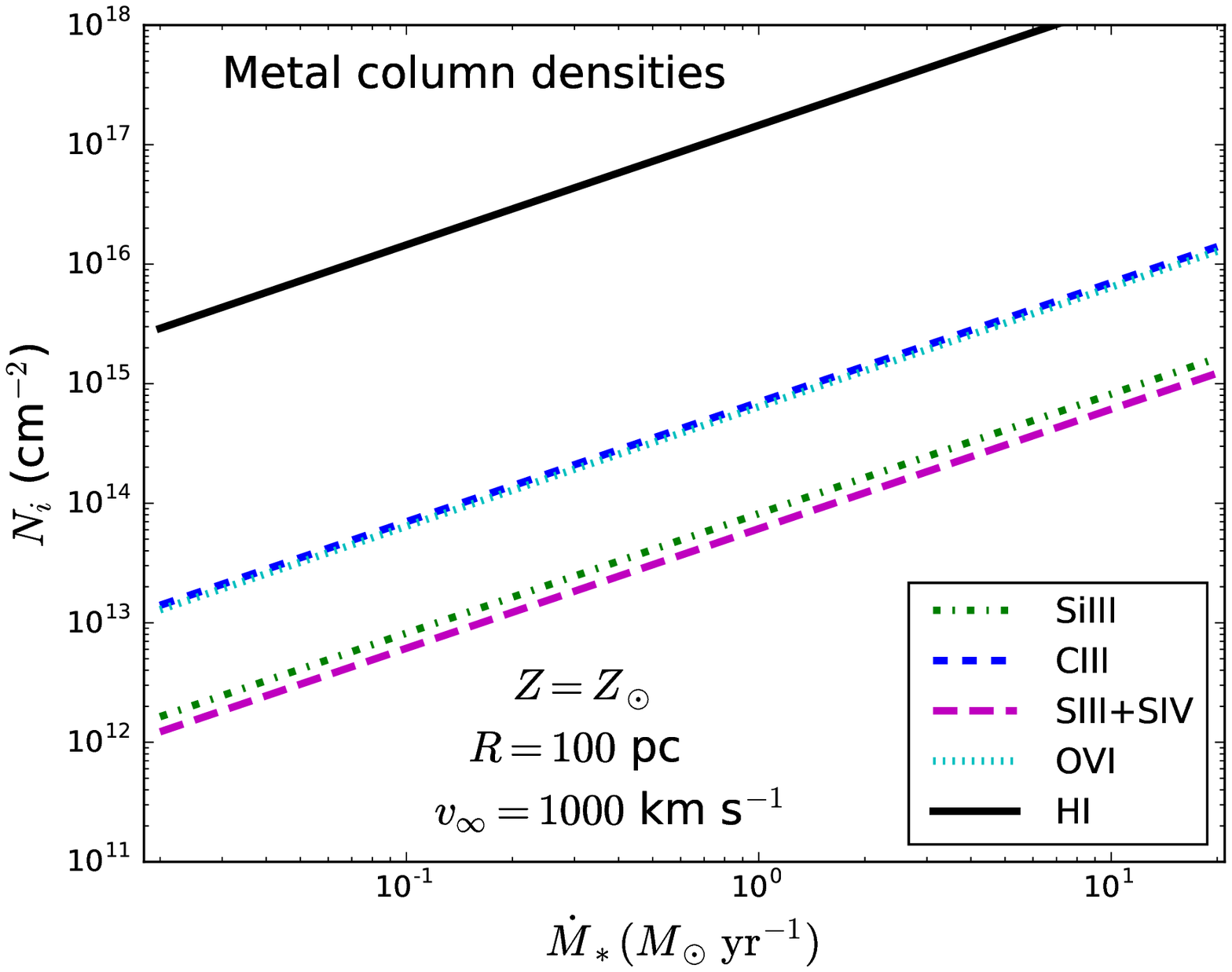}\includegraphics{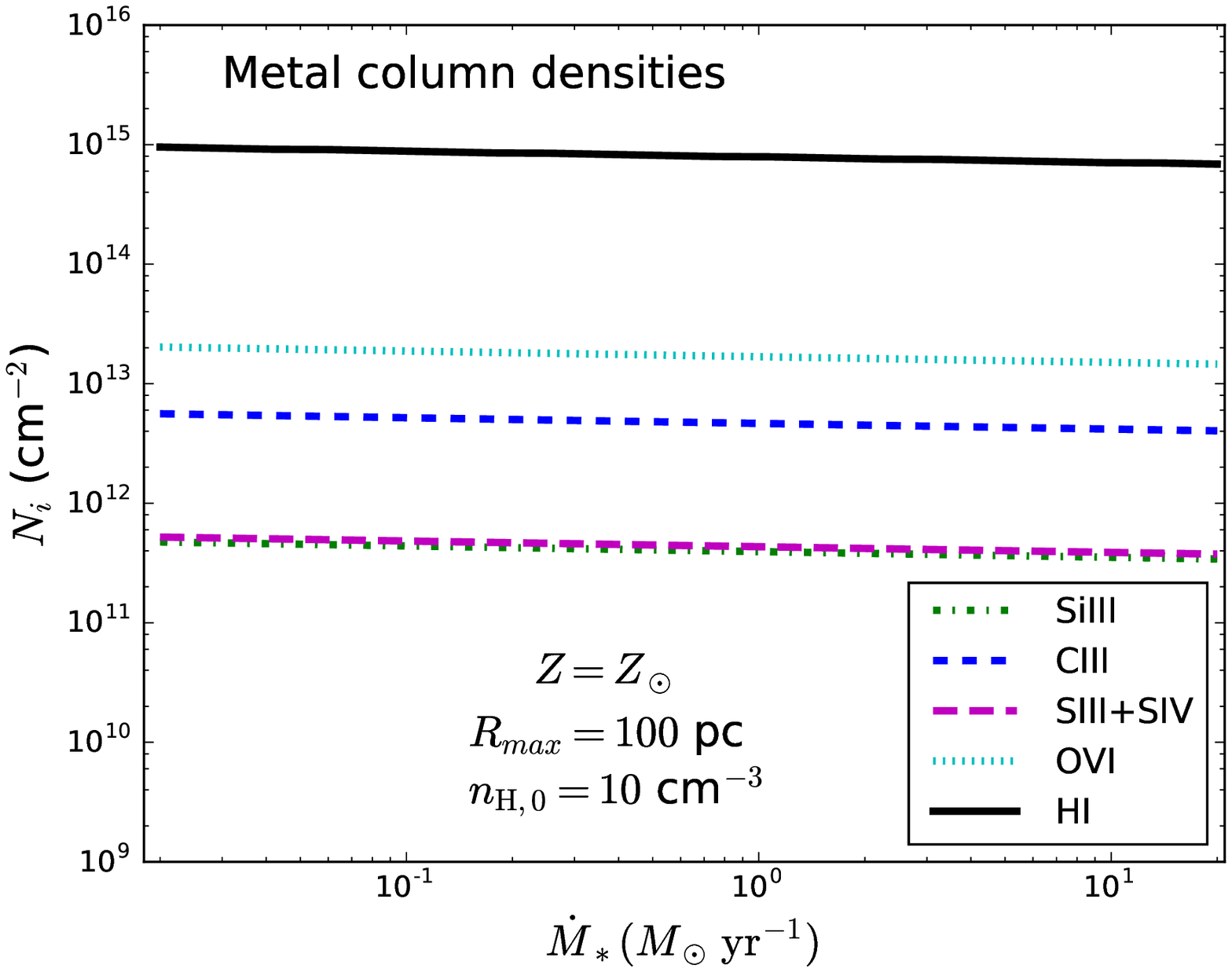}}
\caption{Column densities for selected metal ions, shown for the
  steady state model for $v_\infty=1000\kms$ (left panel) and the
  superbubble model for $n_{\rm H,0}=10\,{\rm cm}^{-3}$ (right
  panel). Also shown is the \HI\ column density. A source region of
  100~pc radius is adopted for the steady state model, and a maximum
  bubble radius of 100~pc for the superbubble model. Solar metallicity
  is assumed for both.
}
\label{fig:MetalsSSTC}
\end{figure*}
 
For the steady-state wind model with $v_\infty>500\kms$ and wind core
radius $R=100$~pc, the temperatures at which these ion abundances peak
are achieved only outside the core, where the gas density is rapidly
declining.
The column density of a typical ion like
$\CIII$ is negligibly small, $N_{\rm CIII}\simeq10^7\,{\rm
  cm^{-2}}$. Of the ions listed above, only \OVI\ would achieve a
measurable column density within the core, $N_{\rm
  OVI}\simeq10^{14}\,{\rm cm^{-2}}$. It is noted, however, that at the
interface of the wind shock and the interstellar medium in the
galactic disc, detectable levels of absorption may arise
\citep{1996ApJS..102..161D}. Such systems could possibly be
distinguished from those produced by superbubbles, discussed below,
through their kinematics.

For $r\gg R$, the wind temperature is
$T\simeq 6.7v_\infty^2(R/r)^{4/3}$. For a given ion $i$, the radius
and density at which $T=T_{\rm max}=10^5\,{\rm K}\,T_{\rm max,5}$ are
$r_{\rm max}\simeq23 R T_{\rm max, 5}^{-3/4}(v_\infty/1000\kms)^{3/2}$
and
$n_{\rm H, max}\simeq4.4\times10^{-4}\,{\rm cm^{-3}}T^{3/2}_{\rm max,
  5}(v_\infty/1000\kms)^{-6}\epsilon\nu_{100}{\dot M_*}R_{100}^{-2}.$
The corresponding column density will be
\begin{eqnarray}
N_i&\simeq& n_{\rm H,max}r_{\rm max}\xi_{i, {\rm max}}\nonumber\\
&\simeq&5.7\times10^{14}\,{\rm  
  cm^{-2}}\epsilon\nu_{100}{\dot  
  M_*}R_{100}^{-1}\left(\frac{1000\kms}{v_\infty}\right)^{9/2}\nonumber\\
&&\times\left(\xi_{i,{\rm max}}T_{\rm max}^{3/4}\right).  
\label{eq:coldenSS}
\end{eqnarray}  
The column densities decline very rapidly with
$v_\infty$. Representative values for some common ions, computed
numerically by integrating along the full wind solution at $r>R$, are
shown in Fig.~\ref{fig:MetalsSSTC} (left panel). These represent the
maximum column densities that would arise from the homogeneous wind
for lines of site passing through the region $r\simeq r_{\rm max}$. At
larger distances, the column densities will rapidly decline.

In the superbubble model, the column densities are dominated by
absorption from a very thin layer at the bubble interface with the
interstellar medium. For the scale-height limited case with a bubble
radius $R_{\rm B}$, the characteristic temperature scale height at the
position where $T=T_{\rm max}$ for a given ion is $L_T=\vert dr/d\log
T\vert_{\rm max}=(5/2)R_{\rm B}(T_{\rm
  max}/T_c)^{5/2}\simeq5.5\times10^{-7}R_{\rm B}(f_T R_{\rm B,
  100}/\epsilon\nu_{100}\dot M_*)^{5/7}T_{\rm max,5}^{5/2}$,
corresponding to a hydrogen density $n_{\rm H, max}\simeq110\,{\rm
  cm^{-3}}\,n_{\rm H,0}^{1/3}{\epsilon\nu_{100}\dot M_*}^{2/3}R_{\rm
  B, 100}^{-4/3}T_{\rm max,5}^{-1}$.  The column density of ion $i$ is
then
\begin{eqnarray}
N_i&\simeq& n_{\rm H, max}\xi_{i, {\rm max}} L_T \label{eq:coldenTC}\nonumber\\
&\simeq& 5.7\times10^8\,{\rm cm^{-2}}\,n_{\rm H,0}^{1/3}(\epsilon\nu_{100}{\dot
  M_*})^{-1/21}R_{\rm B, 100}^{8/21}f_T^{5/7}\nonumber\\
&&\times\left(\xi_{i, {\rm max}}T_{\rm max}^{3/2}\right),
\end{eqnarray}
nearly independent of the star formation rate and only weakly
sensitive to the ambient gas density and size of the
bubble. Representative values, computed numerically by integrating
through the centre of the region $r<R_{\rm B}$, are shown in
Fig.~\ref{fig:MetalsSSTC} (right panel). This is a minimal value that
must arise in the thermal conductive interface with the interstellar
medium. Substantially lower values would be evidence against the
superbubble model. The column density for \HI\ is also a minimal
value, and it will generally be small compared with the \HI\ column
density through the surrounding disc.

\section{Discussion}
\label{sec:Discussion}

\subsection{X-ray constraints}

\begin{figure*}
\scalebox{0.45}{\includegraphics{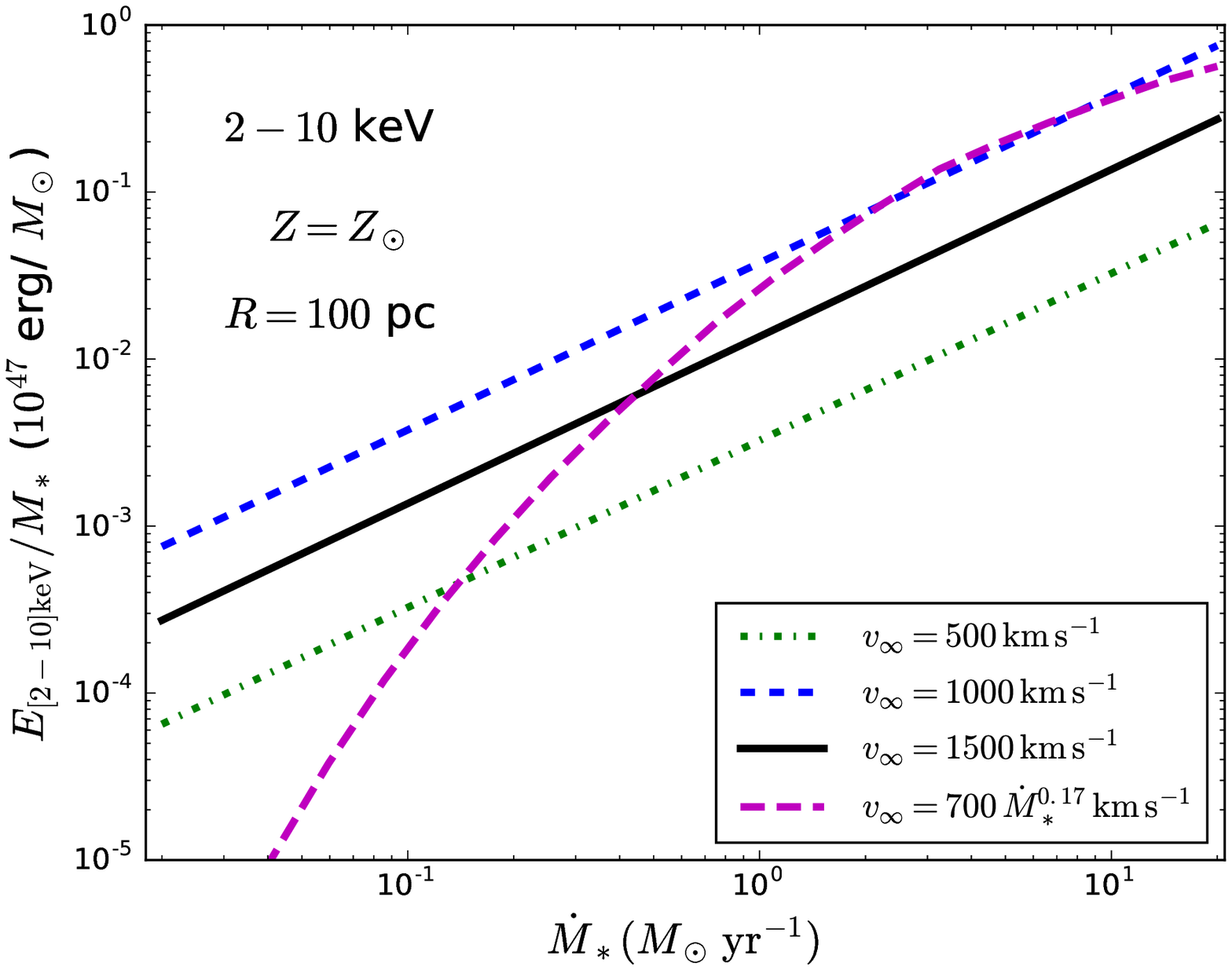}\includegraphics{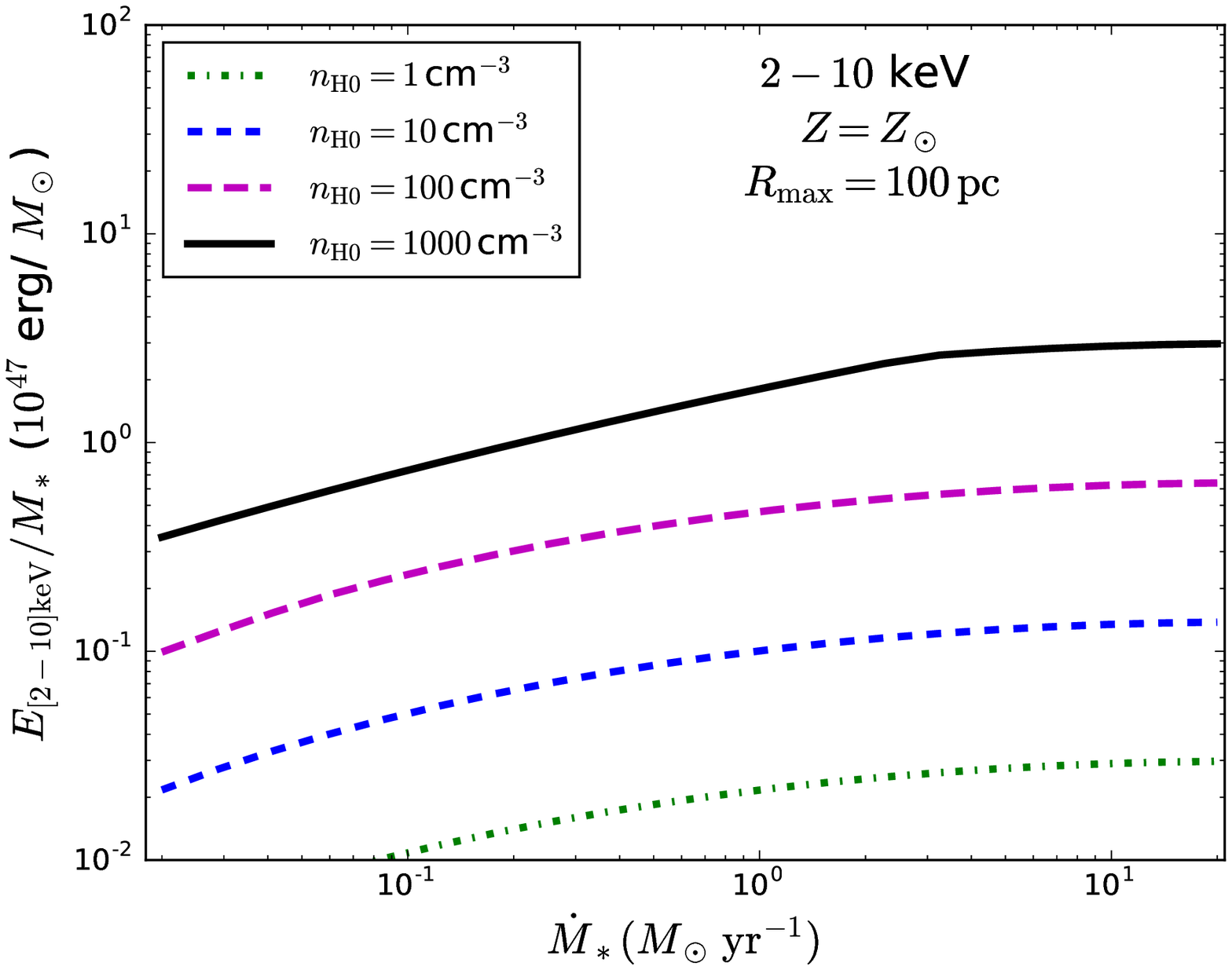}}
\caption{X-ray emission in the 2--10~keV band per solar mass of stars
  formed for a steady-state wind (left panel) and for a superbubble
  (right panel), both as a function of the star formation rate. A
  source region of 100~pc radius is adopted for the steady state
  model, and a maximum wind radius of 100~pc for the
  superbubble. Solar metallicity is assumed for both models.
}
\label{fig:EmissSSWTC}
\end{figure*}

Both the freely expanding steady-state wind model and the
self-similarly expanding superbubble model with thermal heat
conduction may account for the measured amount of diffuse soft x-ray
energy generated per unit mass in stars formed. An additional
assumption of a tight correlation between the asymptotic wind velocity
and the star formation rate, however, is required for the steady state
model. By contrast, thermal evaporation from the cavity walls in the
superbubble model naturally accounts for the measured amount of difuse
x-ray energy per unit mass in stars formed for characteristic
interstellar gas densities $1<n_{\rm H,0}<100\,{\rm cm^{-3}}$, with
the higher values favoured if much of the soft x-ray emission is
absorbed internally to the galaxies.

In the steady state model, the required correlation between the
asymptotic wind velocity and the star formation rate, particularly for
the high x-ray luminosities when internal galactic absorption is
allowed for, is close to the minimum wind velocity
(Eq.~[\ref{eq:vinfmin}]) for which a steady-state wind may be
maintained against radiative cooling within the star-forming region,
consistent with the narrow range in observed radiative
efficiencies. But it does not provide a reason for the narrow
range. One possibility is that the winds are driven by
superbubbles. Once a superbubble expands to the scale-height of the
galactic disc, its thermal pressure drives a vertical conical outflow
rather than further expansion into the disc
\citep{1985ApJ...299...24S, 1988ApJ...324..776M}. Simulations suggest
the outflow is nearly adiabatic \citep{2015MNRAS.453.3499K}, so that
it may be approximated by the steady state model with a superbubble as
the source. The rate of mechanical energy injected by supernovae and
the rate of evaporative mass loss from the disc may be used to define
an effective asymptotic wind velocity for the superbubble:

\begin{equation}
v_\infty^{\rm eff}=\left(\frac{2\dot E}{\dot M_{\rm 
      ev}}\right)^{1/2}=850\,\kms \left(\epsilon\nu_{100}{\dot M_*}/f_TR_{\rm 
    B, 100}\right)^{1/7},
\label{eq:vinfeff}
\end{equation}
close to the required relation found for the steady-state wind
solution. This may reconcile the two models:\ thermal heat conduction
sets the source terms that initiate the wind, which then \lq blows
out' vertically into a steady-state outflow
\citep{1989ApJ...337..141M, 1999ApJ...513..142M, 2003ApJ...599...50F,
  2004ApJ...613..159F, 2014MNRAS.442.3013K}.

For a bubble to blow out, two criteria must be satisfied:\ the cooling
radius must exceed the disc scale height and the bubble velocity must
exceed the sound speed in the surrounding medium. Both of these may be
expressed as a restriction on the average star formation rate per
superbubble cross-sectional area, $\dot\Sigma_{\rm B}={\dot M_*}/\pi
R_{\rm B}^2$. The cooling criterion gives
\begin{eqnarray}
\dot\Sigma_{\rm B}>\dot\Sigma_{\rm B, R} &\simeq&0.0004\,\msun\,{\rm 
  yr^{-1}\, kpc^{-2}}\,n_{\rm 
  H,0}^{7/4}R_{\rm B, 100}^{3/4}\nonumber\\
&&\times(\epsilon\nu_{100})^{-11/4}f_T^{15/8}(\zeta_m+0.15)^{21/8}.
\label{eq:SBR}
\end{eqnarray}
The hydrodynamical computations of \citet{1988ApJ...324..776M} suggest
for the dynamical criterion that, in terms of their dynamical variable
$D$, the bubble velocity must exceed the disc sound speed by a factor
$D^{1/3}\simeq4.6$, for $D>100$. This gives
\begin{equation}
\dot\Sigma_{\rm B}>\dot\Sigma_{\rm B, R} \simeq0.003\,\msun\,{\rm 
  yr^{-1}\, kpc^{-2}}\,T_{\rm d,4}^{3/2}\frac{n_{\rm H,
    0}}{\epsilon\nu_{100}},
\label{eq:SBD}
\end{equation}
where $T_{\rm d,4}$ is the temperature of the ambient disc gas in
units of $10^4$~K. The dynamical criterion is similar to the estimate
of \citet{2004ApJ...606..829S}. For
$n_{\rm H, 0}\sim1-10\,{\rm cm^{-3}}$, these criteria give
$\dot\Sigma_{\rm B}>0.003-0.03\,\msun\,{\rm yr^{-1}\, kpc^{-2}}$,
comparable to the minimum observed star formation surface density in
galaxies with winds \citep{2005ARAA..43..769V}. This raises the
question:\ do the proxies for star formation actually probe the larger
region of a superbubble, so that the minimum observed star formation
surface density in galaxies with winds may be identified with the
minimum $\dot\Sigma_{\rm B}$ required for blowout?

Differences in the hard x-ray luminosities of the wind cores are
expected between the two models. In the steady-state model, the
exponential sensitivity to the star formation rate results in a rapid
decrease with decreasing star formation rate of the 2--10~keV
luminosity when the correlation $v_\infty\simeq700{\dot M_*}^{1/6}$ is
imposed, as shown in Fig.~\ref{fig:EmissSSWTC}. At
$E_{[2-10]{\rm keV}}/M_*\simeq2\times10^{45}\,{\rm erg\,M_\odot^{-1}}$
for $\dot M_*=1\,{\rm M_{\odot}\,yr^{-1}}$, the specific emissivity is
up to an order of magnitude smaller than the predicted value for a
superbubble with $10<n_{\rm H,0}<100\,{\rm cm^{-3}}$. For
$M_*=0.1\,{\rm M_{\odot}\,yr^{-1}}$, at
$E_{[2-10]{\rm keV}}/M_*\simeq2\times10^{43}\,{\rm erg\,M_\odot^{-1}}$
the predicted specific emissivity in the steady state model is nearly
two orders of magnitude smaller than for a superbubble with
$n_{\rm H,0}\simeq1\,{\rm cm^{-3}}$. The hard x-ray luminosity of the
wind cores may thus serve to discriminate between the models.

Both models have been scaled to a supernova ejecta mechanical energy
of $10^{51}\epsilon\,{\rm erg}$. Observations suggest a range of
$0.6<\epsilon<1.5$ \citep{1989ARA&A..27..629A}. The effects on the
values of $v_\infty$ in the steady state model or $n_{\rm H,0}$ in the
superbubble model required to match observations may be inferred from
the scaling relations in Eqs~(\ref{eq:ExMsssw}) and
(\ref{eq:ExMssswtc}). This is a systematic uncertainty of the models,
and applies as well to the predictions for radio power and metal
column densities.

\subsection{Radio constraints}

The radio spectrum in the steady state model is expected to steepen
due to the depletion of relativistic electrons by synchrotron and
plasmon losses over a timescale of $\sim10$~Myr. The superbubble model
is computed only until the bubble emerges from the galactic disc,
which is typically a much shorter time. An assessment of the long term
energy losses of the relativistic electron population may be
approximated by imposing energy losses over a characteristic age of
10~Myr on the bubble. Doing so shows that, because of the low gas
density inside the bubble cavity, the relativistic electron population
is not depleted by synchrotron or plasmon losses, so that the emission
persists without steepening. Although the intrinsic electron
distribution may provide a steep spectrum, this is not required. A
hard spectrum near 1.4~GHz may then provide a means of discriminating
between the superbubble model and the conduction-free steady state
model.

While both the steady state and the superbubble models are consistent
with the measured radio powers for plausible values of the synchrotron
model parameters, the data do not require these values. Other
combinations will match the data. It is also possible other
mechanisms, such as large-scale cosmic ray emission, dominate the
measured power if $f_e\ll1$. Radio measurements that resolve the
star-forming regions in starbursts may help to clarify the
contribution of shock-accelerated electrons within the wind-generating
region to the total synchrotron radio emission.

\subsection{Metal column density constraints}

\begin{figure}
\scalebox{0.45}{\includegraphics{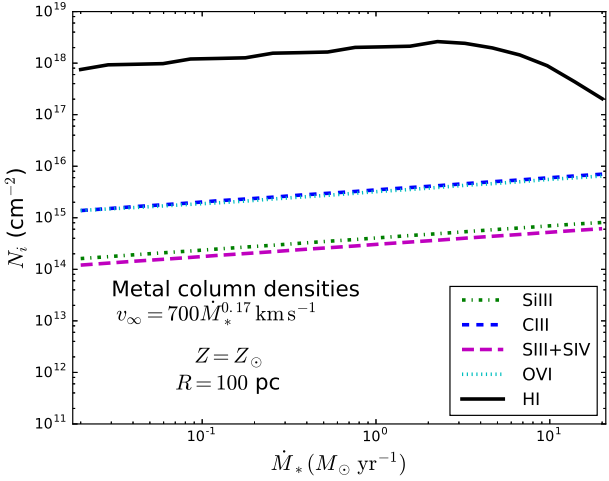}}
\caption{Column densities for selected metal ions, shown for the
  steady state model with $v_\infty=700 {\dot M_*}^{0.17}\kms$. Also
  shown is the \HI\ column density. A source region of 100~pc radius
  and solar metallicity are adopted.
}
\label{fig:MetalsSSC}
\end{figure}

\begin{figure}
\scalebox{0.45}{\includegraphics{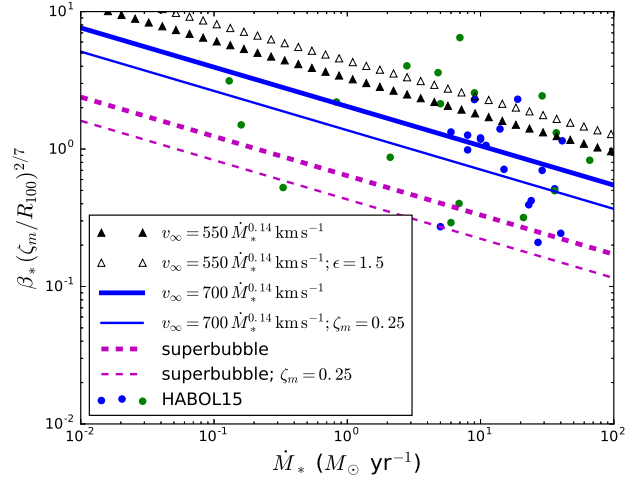}}
\caption{Mass loading factor $\beta_*(\zeta_m/R_{100})^{2/7}$ as a
  function of star formation rate. To allow for variations in the size $R=100R_{100}$~pc of the star-forming
  regions and the gas metallicity $\zeta_m$, the mass loading factor has
  been scaled by the factor $(\zeta_m/R_{100})^{2/7}$ (see text). The
  upper limit imposed by radiative cooling in the freely-expanding
  steady state model is shown for $\epsilon=1$ (filled triangles) and
  $\epsilon=1.5$ (open triangles). The predicted value in the steady
  state model for $v_\infty=700\kms(\epsilon^2\nu^2_{100}\dot
  M_*/R_{100})^{1/7}$ is shown for $\zeta_m=1$ (thick blue solid line)
  and $\zeta_m=0.25$ (thin blue solid line). The prediction for the
  superbubble model when the bubble reaches size $R$ is shown for
  $\zeta_m=1$ (thick magenta broken line) and $\zeta_m=0.25$ (thin
  magental broken line). The data are from
  \citet{2015ApJ...809..147H}:\ the dark (blue) circles represent
  strong cloud outflows and light (green) circles represent weak or no
  cloud outflows.
}
\label{fig:betas}
\end{figure}

Imposing the approximate analytic velocity correlation
$v_\infty\simeq700\kms(\epsilon^2\nu_{100}^2\dot M_*/R_{100})^{1/7}$
(Sec.~\ref{subsec:numssw_xray}) for the steady state model results in
ion column densities depending only weakly on the star formation rate
and size of the wind-generating region,
\begin{eqnarray}
N_i&\simeq&2.8\times10^{15}\,{\rm 
  cm^{-2}}(\epsilon\nu_{100})^{-2/7}\nu_{100}({\dot 
  M_*}/R_{100})^{5/14}\nonumber\\
&&\times\left(\xi_{i,{\rm max}}T_{\rm max}^{3/4}\right). 
\label{eq:coldenSSv700}
\end{eqnarray} 
The metal column densities computed after imposing the similar
numerical correlation $v_\infty=700\kms{\dot M_*}^{1/6}$ on the
numerically integrated model with $R=100$~pc are shown in
Fig.~\ref{fig:MetalsSSC}. Commonly detected ions will have column
densities between $10^{14}-10^{16}\,{\rm cm^{-2}}$, comparable to
those measured \citep{2001ApJ...554.1021H}. The ion abundances peak,
however, at characteristic temperatures near the peak in the cooling
curve, so that the gas will be thermally unstable. The computed column
densities would then represent minimal values:\ as the gas condenses,
the column densities will increase, although the covering fraction of
the absorption systems will decline as the gas fragments. Measurements
suggest the column densities of measured systems are determined by the
velocity at which they flow through coronal temperatures
\citep{1986ApJ...310L..27E, 2002ApJ...577..691H}, for which the column
density will be of the order
$(1/\sigma_i){\mathcal M}_{\rm
  cool}$\footnote{\citet{1986ApJ...310L..27E}
  and \citet{2002ApJ...577..691H} argue the cooling column density is
  $3kT v_{\rm cool}/\Lambda(T)$, where $v_{\rm cool}$ is the flow
  velocity of the cooling gas and
  $\Lambda(T)\sim E_i\sigma_i(T) v_e(T)$ is the cooling coefficient
  due to collisional excitation of a transition with energy $E_i$, and
  $v_e$ is the velocity of the electrons. Coronal temperatures
  correspond to $k_{\rm B}T\sim E_i$.}, where $\sigma_i(T)$ is the
characteristic collisional cross section of the ion and
${\mathcal M}_{\rm cool}(T)$ is the Mach number of the flow.

The neutral hydrogen column density arises from a recombination layer
at $T\sim10^4$~K that will occur at radial distances
$r\simeq76R(\epsilon\nu_{100})^{3/7}(\dot M_*/R_{100})^{3/14}$, typically
5--15~kpc. The layer will be optically thick to any penetrating
external photoionizing radiation that may modify the predicted metal
column densities, although they would still be susceptible to
photoionizing radiation from the starburst. Thermal instabilities,
however, may cause the layer to fragment so that its structure may
become porous.

The superbubble model predicts metal column densities much smaller
than measured. This was recognised by \citet{2001ApJ...554.1021H}, who
suggested dynamical instabilities as the superbubble emerges from the
disc will develop turbulent mixing layers that, in sufficient number,
could produce the measured column densities.

Measurements of absorption lines produced by clouds in wind outflows
have been used by \citet{2015ApJ...809..147H} to assess the amount of
mass loading in the winds.\footnote{The velocities of the clouds
  determined from the widths of their absorption features are not
  necessarily to be identified with the outflow velocity of the wind
  gas impinging on them; the clouds will generally be accelerated by
  ram pressure to lower velocities than the flow because of their
  higher densities \citep[eg][]{2015ApJ...805..158S}.} Their results
are shown in Fig.~\ref{fig:betas} along with the predictions of the
models. The measured values have been rescaled to account for the
variation in the star-forming region sizes and the
metallicities. Following the terminology of these authors, the points
have also been designated as either corresponding to a strong outflow
or to a weak or no outflow.  The upper values lie close to the maximum
allowed by cooling in the steady state model, and include many of the
weak or no outflows. Adopting the correlation
$v_\infty=700\kms (\epsilon^2\nu_{100}^2\dot M_*/R_{100})^{1/7}$
results in values matching the median of those measured, corresponding
to strong outflows.

The lowest mass loading values lie near the value for the superbubble
model, and correspond to many of the weak or no outflows. (Here, the
measured star formation region is identified with the superbubble
radius.) This supports the possibility that the steady-state wind is
initiated by a superbubble. The high metal column densities may then
arise within an outflow region at coronal temperatures, as for the
steady-state wind. Further mass loading from the hydrodynamical
ablation or thermal evaporation of clouds above the plane would
account for the larger mass loading factors. Observational evidence
suggests thermal conduction is active within halo clouds
\citep{2005MNRAS.362..626M}. The clouds may have pre-existed or been
accreted from larger scales, as simulations suggest entrained clouds
have too short survival times to have originated in the disc
\citep{1994ApJ...420..213K, 2015ApJ...805..158S} unless possibly
supported by magnetic field pressure \citep{1994ApJ...433..757M,
  2015MNRAS.449....2M}.

\section{Conclusions}
\label{sec:conclusions}

Two models for the generation of galactic winds are compared. While
both model the winds as powered by supernovae explosions in active
areas of star formation, the models differ fundamentally in their
microphysical assumptions. One neglects thermal heat conduction and
treats the wind in a steady state of free expansion, as may be
expected for an age large compared with the flow time. The second
includes thermal heat conduction, resulting in time-dependent
self-similar expansion into a surrounding medium. Observational
predictions of the models are compared with x-ray, radio and UV metal
line absorption measurements. The principal conclusions are summarised
below.

\subsection{Steady-state wind}
\label{subsec:ssm}

The steady state model is characterised by two principal parameters,
chosen here to be the star formation rate $\dot M_*$ and the
asymptotic wind velocity $v_\infty$. The wind is assumed to be
generated by uniformly distributed supernovae within a core region of
radius $R=100R_{100}$~pc. The requirement that the energy generation
rate exceed the cooling rate imposes a lower limit on the asymptotic
velocity of
$v_\infty>550\,{\rm km\,s^{-1}}(\epsilon\nu_{100}{\dot M_*}\zeta_m
R_{100}^{-1})^{0.14}$,
where $\zeta_m$ is the metallicity of the wind gas relative to solar
and $\nu_{100}$ and $\epsilon$ are scaling parameters describing the
expected supernovae rate per star formed and their amount of energy
injection. This restriction constrains the mass-loading factor to
$\beta_*<3.3(\epsilon\nu_{100})^{0.73}(\dot
M_*\zeta_mR_{100}^{-1})^{-0.27}$.

The amount of soft x-ray energy generated per unit mass of stars
formed scales approximately as $E_x/M_*\sim {\dot M_*}/
(v_\infty^{1/\alpha} R)$ with $\alpha=1/7-1/5$. This is contrary to
the measurements of \citet{2012MNRAS.426.1870M}, who find a constant
amount of soft x-ray energy generation per unit mass of stars formed,
independent of the star formation rate. The amount measured in the
0.5--2~keV band of $2\times10^{46}\,{\rm erg\, M_\odot^{-1}}$ is
matched if the asymptotic wind velocity correlates with the star
formation rate according to $v_\infty\simeq 1000\kms
(\epsilon^2\nu_{100}^2\dot M_*/R_{100})^\alpha$ with
$\alpha\sim1/7-1/5$. A lower asymptotic speed is required if the
measured x-ray luminosities are reduced by extinction internal to the
galaxies. Matching to the extinction-corrected 0.3--10~keV band value
of $2\times10^{47}\,{\rm erg\, M_\odot^{-1}}$ requires $v_\infty\simeq
700\kms (\epsilon^2\nu_{100}^2\dot M_*/R_{100})^\alpha$ with
$\alpha\sim1/7-1/5$, close to the correlation required to sustain a
steady-state wind against radiative cooling and suggesting only a
narrow range of mass loading is allowed for a given star formation
rate. The predicted amount of hard x-ray energy generated in the
2--10~keV band declines steeply with decreasing star formation rate,
falling from $\sim2\times10^{45}\,{\rm erg\, M_\odot^{-1}}$ at $\dot
M_*=1\,{\rm M_\odot\,yr^{-1}}$ to $\sim2\times10^{43}\,{\rm erg\,
  M_\odot^{-1}}$ at $\dot M_*=0.1\,{\rm M_\odot\,yr^{-1}}$.

The model gives a near constant amount of radio synchrotron energy at
1.4~GHz generated per unit mass of stars formed. The value is
sensitive to the assumed parameters of the energy distribution
function of relativistic electrons, but agreement with the data may be
achieved for typical values inferred from supernova remnants. The
spectrum, however, is expected to be steeper than would be given by
the electron energy index for a typical wind age of 10~Myr because of
the depletion of relativistic electrons within the wind-generating
core by synchrotron and plasmon wave energy losses.

Except for high ionization species like \OVI, commonly measured metal
absortion lines such as \CIII, \SiIII, \SIII\ and \SIV are expected to
be undetectable within the hot core of the wind. (Since the wind core
resides within the galactic disc, these ions may form in detectable
amounts in the disc within the interface between the wind shock front
and the interstellar medium.) Detectable levels of these ions are
expected from the outflow region with typical column densities of
$10^{14}-10^{16}\,{\rm cm^{-2}}$, weakly increasing with the star
formation rate after allowing for the correlation between the
asymptotic wind velocity and the star formation rate required to match
the soft x-ray data. Since the gas is likely thermally and dynamically
unstable in the temperature regime at which the ions form in greatest
abundance, these values may set lower limits on the column densities
of an increasingly fragmenting outflow.

Estimates of the amount of mass-loading from absorption line data show
a maximum mass-loading factor comparable to the maximum imposed by the
cooling restriction. For the approximate analytic velocity correlation
$v_\infty\simeq 700\kms (\epsilon^2\nu_{100}^2\dot
M_*/R_{100})^{1/7}$, the mass loading factor is somewhat smaller than
the cooling limit,
$\beta_*\simeq2.0(\epsilon\nu_{100})^{3/7}(R_{100}/\dot M_*)^{2/7}$,
comparable to measured values and corresponding to a mass injection
rate into the wind of $\dot M\simeq2.0\msun\,{\rm
  yr}^{-1}\,(\epsilon\nu_{100})^{3/7}{\dot M_*}^{5/7}R_{100}^{2/7}$.

\subsection{Superbubble with thermal heat conduction}
\label{subsec:sbmtc}

The superbubble model may be parametrized by the star formation rate
and the ambient hydrogen density into which the bubble expands within
a disc galaxy. The model ceases to be directly applicable once the
bubble breaks out of the disc:\ hydrodynamical simulations have shown
instead the bubble vents its thermal energy into conical winds
vertical to the disc. This will occur for expected hydrogen densities
within the discs before the bubble growth is restricted by the onset
of radiative cooling. The subsequent evolution of the wind is not
directly computed here.

Mass loading by thermal evaporation naturally results in an amount of
soft x-ray generation per solar mass of stars formed only weakly
dependent on the star formation rate. Matching to the measured value
in the 0.5--2~keV band requires ambient hydrogen densities of
$1-10\,{\rm cm^{-3}}$, while higher densities of
$10-100\,{\rm cm^{-3}}$ are required to match the measured value in
the 0.3--10~keV band, after allowing for extinction corrections for
internal galactic absorption. (Since extended emission out of the disc
will contribute a comparable amount to the measured x-ray luminosity,
the required densities may be somewhat lower.) For these densities,
cooling and dynamical criteria require the star formation rate per
superbubble cross-sectional area to exceed
$0.003-0.03\,\msun\,{\rm yr^{-1}\,kpc^{-2}}$. A higher level of hard
x-ray emission in the 2--10~keV band is expected for a superbubble
compared with the wind core in the steady-state wind model for low
star formation rates, with a value in excess of
$\sim10^{45}\,{\rm erg\, M_\odot^{-1}}$, and possibly as much as an
order of magnitude larger or more, sustained by the superbubble for
star formation rates $\dot M_*\sim0.1\,{\rm M_\odot\,yr^{-1}}$. Such
high radiative efficiencies measured in the wind cores would favour
the superbubble model.

If the bubble blows out into the halo, it will do so with a thermal
pressure corresponding to an effective asymptotic wind velocity, in
terms of the steady-state wind parametrization, of $v_\infty^{\rm
  eff}\simeq 850\,\kms (\epsilon\nu_{100}{\dot M_*}/f_TR_{\rm B,
  100})^{1/7}$.  The effective wind velocity is suggestively close to
the value required for the steady state model to match the soft x-ray
data. This may indicate that large-scale winds are indeed driven by
superbubbles. The amount of mass loading through thermal evaporation,
however, is somewhat
low:\ $\beta_*\simeq0.64(\epsilon\nu_{100})^{5/7}(R_{\rm B,
  100}f_T/{\dot M_*})^{2/7}$, corresponding to a mass injection rate
of $\dot M\simeq0.64\msun\,{\rm yr}^{-1}(\epsilon\nu_{100}\dot
M_*)^{5/7}(R_{\rm B, 100}f_T)^{2/7}$, just somewhat smaller than the
amount above for the steady state model to match the amount of x-ray
energy generated per unit mass of stars formed. The value for mass
loading does correspond to, and so may set, the minimal values
measured from metal absorption line data. The larger measured values,
however, would require additional mass loading as the wind left the
disc, such as from hydrodynamical ablation or the thermal evaporation
of clouds, for which there is observational evidence.

As for the steady state model, the measured 1.4~GHz radio luminosities
of galaxies with winds may be recovered for plausible parameters of
the relativistic electron energy distribution. Unlike for the steady
state model, the relativistic electron distribution is not depleted
within the wind core so that steep radio spectra are not
required. This provides a possible means of distinguishing between the
models.

Detectable levels of \SiIII, \CIII, \SIII\, \SIV and \OVI in
absorption are expected through the superbubble, with typical column
densities of $10^{11}-10^{14}\,{\rm cm^{-2}}$, nearly independent of
the star formation rate and only weakly dependent on the ambient gas
density and size of the superbubble. Although an ionization layer
produced by a wind shock impacting on the interstellar medium of a
galaxy in the steady state model may produce detectable levels of
metal absorption as well, differences in the column density ratios and
kinematics between the models would be expected. It is emphasized that
the thermal evaporation layer in the superbubble model sets a lower
limit to the metal absorption, which should have a covering fraction
of order unity. The absence of the minimal predicted levels of
absorption would be a telling indicator against the superbubble
model. Larger values may also arise outside the disc in gas at coronal
temperatures in a superbubble-driven outflow.



\appendix

\section{Steady-state wind}
\label{app:SSwind}

The equations governing a spherical wind driven by stellar winds and
supernovae are:

\begin{equation}
\frac{\partial\rho}{\partial t}+\frac{1}{r^2}\frac{d}{dr}\left(\rho v r^2\right) = \alpha\rho_*,
\label{eq:mass}
\end{equation}

\begin{equation}
\rho\frac{\partial v}{\partial t}+\rho
v\frac{dv}{dr}=-\frac{dP}{dr}-\rho g-\alpha\rho_* v,
\label{eq:momentum}
\end{equation}

\noindent and

\begin{equation}
  \frac{\partial}{\partial t}\left(\frac{1}{2}\rho v^2+\rho\epsilon\right)+\frac{1}{r^2}\frac{d}{dr}\left[\rho v
    r^2\left(\frac{1}{2}v^2+\frac{\gamma}{\gamma-1}\frac{P}{\rho}\right)\right]=\alpha\rho_*\langle q\rangle,
\label{eq:energy}
\end{equation}

\noindent where $v$, $\rho$, $P$ and $\epsilon$ are the gas radial
velocity, density, pressure and thermal energy per unit mass,
respectively, $\gamma$ is the ratio of specific heats, $\rho_*$ is the
density distribution of stars, $\alpha$ is the combined rate of mass
loss by stars and supernovae per unit mass in stars, and
$\langle q\rangle$ is the average energy injection rate by the stars
and supernovae per unit mass \citep{1968MNRAS.140..241B,
  1971ApJ...165..381J, 1971ApJ...170..241M, 1985Natur.317...44C}. For
a simple wind model with the sources driving the wind confined to a
region $r<R$, with $\alpha\rho_*$ and $\langle q\rangle$ constant
within $r<R$ and vanishing outside, it is convenient to define the
dimensionless quantities $x=r/R$, $\tilde v=v/\langle q\rangle^{1/2}$,
$\tilde\rho=\rho/(\alpha\rho_*\langle q\rangle^{-1/2}VR^{-2})$ and
$\tilde P=P/(\alpha\rho_*\langle q\rangle^{1/2}VR^{-2})$, where
$V=(4\pi/3)R^3$. In terms of the energy injection efficiency
$\epsilon$ and mass loading factor $\beta$ of Sec.~\ref{subsec:ssw},
the total energy and mass injection rates are then
$\dot E=\epsilon {\dot E}_1=\alpha\rho_* V\langle q\rangle$ and
$\dot M=\beta\dot M_1=\alpha\rho_* V$. Here $\dot E_1$ is given by
an injection of $10^{51}\nu_{100}$~erg of mechanical energy per
$100M_\odot$ of stars formed along with mass injected at the rate
$\dot M_1$. In terms of the asymptotic wind velocity $v_\infty$,
$\langle q\rangle=v_\infty^2/2$.

For a wind in a steady state with negligible gravity, the velocity and
pressure at $x<1$ may then be expressed in terms of the density as

\begin{equation}
\tilde v(x) = \frac{x}{4\pi\tilde\rho(x)}
\label{eq:vrho}
\end{equation}

\noindent and

\begin{equation}
\tilde P(x) = \frac{\gamma-1}{\gamma}\tilde\rho(x)\left[1-\frac{1}{2}\left(\frac{x}{4\pi\tilde\rho(x)}\right)^2\right].
\label{eq:Prho}
\end{equation}

\noindent A smooth transition from subsonic flow near the centre to
supersonic flow at large $x$ requires the Mach number
$M=v/c_{\rm ad}$, where $c_{\rm ad}=(\gamma P/\rho)^{1/2}$ is the
adiabatic sound speed, to satisfy $M=1$ at $x=1$, giving

\begin{eqnarray}
  \left(\frac{3\gamma+1/M^2}{1+3\gamma}\right)^{-(3\gamma+1)/(5\gamma+1)}&\left(\frac{\gamma-1+2/M^2}{1+\gamma}\right)^{(\gamma+1)/[2(5\gamma+1)]}\\
&=x\qquad (x<1)
\label{eq:Mach_in}
\end{eqnarray}

\noindent and

\begin{eqnarray}
M^{2/(\gamma-1)}&\left(\frac{\gamma-1+2/M^2}{1+\gamma}\right)^{(\gamma+1)/[2(\gamma-1)]}\\
&=x^2\qquad(x>1)
\label{eq:Mach_out}
\end{eqnarray}

\noindent \citep{1985Natur.317...44C}. These equations may be solved
numerically, and $\tilde\rho$ extracted. For $x<1$, however, the
density is well-approximated by

\begin{equation}
\tilde\rho(x)\simeq\tilde\rho_0\left\{1+\log\left[1-\frac{1}{2}\frac{\gamma+1}{\gamma-1}\left(\frac{x}{4\pi\tilde\rho_0}\right)^2\right]^{(3\gamma+4)/(\gamma+1)}\right\},
\label{eq:rho_in}
\end{equation}
where the central density $\tilde\rho_0=\tilde\rho(0)$ is given by
\begin{equation}
\tilde\rho_0=\frac{\left(3\gamma+1\right)^{(3\gamma+1)/(5\gamma+1)}}{4\pi(\gamma-1)^{1/2}}\left(\frac{2}{\gamma+1}\right)^{(\gamma+1)/[2(5\gamma+1)]}.
\label{eq:rho0}
\end{equation}

For a monatomic gas ($\gamma=5/3$), the central density is
$\tilde\rho_0=(3/2)^{1/2}(3/4)^{1/7}6^{9/14}/4\pi\simeq0.2960$. The
characteristic core radius, corresponding to $\delta(x_c)=-0.5$, is
$x_c\simeq0.98$, so that the density varies little until $x$
approaches 1. The velocity for $x\ll x_c$ is
$\tilde v(x)\simeq0.2689x$ and the central pressure
$\tilde P_0=(2/5)\tilde\rho_0\simeq0.1184$. For $x>1$, the solution is
analytic for $\gamma=5/3$. Defining
$S_\pm(x)=[(4x)^4/2-64\pm8(4x)^2(x^4-1)^{1/2}]^{1/3}$,
$U(x)=[S_+(x)+S_-(x)-4]^{1/2}$ and
$W(x)=\{2(4x)^2/U(x)-[S_+(x)+S_-(x)+8]\}^{1/2}$, the Mach number is
given by $M(x)=[U(x)+W(x)]/2$. Then

\begin{equation}
\tilde v(x)=2^{1/2}\frac{M(x)}{[3+M^2(x)]^{1/2}},
\label{eq:uss}
\end{equation}

\begin{equation}
\tilde\rho(x) =
\frac{1}{2^{1/2}\, 4\pi}\frac{[3+M^2(x)]^{1/2}}{M(x)}\frac{1}{x^2}
\label{eq:rhoss}
\end{equation}

\noindent and

\begin{equation}
\tilde P(x) =
\frac{2^{1/2}\,3}{20\pi}\frac{1}{M(x)[3+M^2(x)]^{1/2}}\frac{1}{x^2}.
\label{eq:Pss}
\end{equation}

\bibliographystyle{mn2e-eprint}
\bibliography{winds}

\label{lastpage}

\end{document}